\documentclass[sn-mathphys-num]{sn-jnl}

\usepackage{graphicx}%
\usepackage{multirow}%
\usepackage{amsmath,amssymb,amsfonts}%
\usepackage{amsthm}%
\usepackage{mathrsfs}%
\usepackage[title]{appendix}%
\usepackage{xcolor}%
\usepackage{textcomp}%
\usepackage{manyfoot}%
\usepackage{booktabs}%
\usepackage{algorithm}%
\usepackage{algorithmicx}%
\usepackage{algpseudocode}%
\usepackage{listings}%

\begin{document}

\title[Article Title]{A cognitive basis for physical time}

\author{\fnm{Per} \sur{\"Ostborn}}

\affil{\orgdiv{Division of Mathematical Physics}, \orgname{Lund University}, \city{Lund}, \postcode{221 00}, \country{Sweden}}

\abstract{
The treatment of time in relativity does not conform to that in quantum theory. In the context of quantum gravity this is called \emph{the problem of time}. A crucial difference is that time $t$ may be seen as an observable in relativity theory, just like position $x$, whereas in quantum theory $t$ is a parameter, in contrast to the observable $x$. Aiming to resolve the discrepancy, a formalization of time in the spirit of Kant's Copernican revolution is suggested, where it is required that the treatment of time in physics agree with our cognition. This leads to reconsideration of the notions of identity and change of objects, as well as the nature of physical states and their evolution. The formalization has two components: sequential time $n$ and relational time $t$. The evolution of physical states is described in terms of $n$, which is updated each time an event occurs. The role of $t$ is to quantify distances between events in space-time. There is a space-time associated with each $n$, in which $t$ represents the knowledge at time $n$ about temporal distances between present and past events. A universal ordering of events in terms of $n$ can be postulated even though distances $t$ are relativistic. In short, it is argued that time as a sequential flow of events should be separated from time as a measure of distance between events. In physical models, these aspects of time can be expressed as one evolution parameter and one observable, respectively.}

\keywords{Problem of time, Kant’s Copernican revolution, Forms of appearances, Sequential time, Relational time, Evolution parameter}

\maketitle

\section{Introduction}
\label{intro}

Philosophers and physicists still struggle with the age-old questions whether the perceived flow of time, and the distinctions between past, present and future, should be granted fundamental significance, or be treated as structure that emerges at a secondary subjective level.

These cognitive aspects of time have become less and less pronounced in physics as time has passed by. It is not necessary to introduce such temporal structure in Newtonian mechanics. However, Newtonian time is clearly separated from the spatial degrees of freedom, in accordance with our subjective perception. This distinction was put into question when special relativity was discovered, since Lorentz transformations between equally valid reference frames mix temporal and spatial coordinates. Further, the relativity of simultaneity made the very concept of a universal present dubious. The birth of general relativity reinforced these conclusions. It made the temporal and spatial intervals between a given pair of events as measured by two observers depend not only on their relative state of motion, but also on their positions in a gravitational field.

These developments seemed to rule out presentism and favor eternalism, a block universe with 'frozen' trajectories of objects without distinctions between past, present and future \cite{crisp}. The removal of this cognitive structure from physical time, and the apparent loss of the special status of time among the degrees of freedom in the formalism, have led number of theorists to take one step further and promote the idea that time should be abandoned altogether as a fundamental concept in physics \cite{barbour,rovelli}. It can be argued that time cannot play any fundamental role since an external clock is needed to measure time \cite{rovelli}. Therefore temporal intervals cannot be defined in the universe as a whole, but only for small parts of the world that are monitored from the outside.

The idea to remove time from physics is gaining traction since no one has yet been able to make the notion of time in general relativity conform with that in quantum theory. In the context of quantum gravity research this is called \emph{the problem of time} \cite{isham,kuchar}. Of course, the simplest way out of this deadlock is to say that there is no time and therefore no problem.

Actually, this is hinted at by a straightforward attempt to express a quantum mechanical evolution equation for the wave function of the entire universe. The result is the Wheeler-DeWitt equation, which has the form of a steady state equation with zero energy, corresponding to a static universe \cite{dewitt}. However, some physicists suspect that this equation is invalid, arguing that quantum mechanics can be applied only to proper subsets of the world, never to the entire universe \cite{smolin}. That would leave room for a dynamic model of the universe as a whole. 

There are more arguments for the continued relevance of time in physics. Even though magnitudes of time intervals between events are not universal in relativity theory, the more primitive notion of time as an ordering of events does retain its meaning. It is built into relativity theory, since the metric has a fixed signature in which one of the four axes of space-time is assigned the opposite sign as compared to the other three. The trajectories of all massive objects are constrained to move within the local light cones in a given direction along this particular axis, whereas they may wiggle back and forth along the other three axes. This feature corresponds to the flow of time, making it possible to order events along any given world line in a linear sequence. As emphasized already by Eddington, relativity makes an absolute distinction between time and space in the sense that the relation between a pair of events is either time-like or space-like. "It is not a distinction between time and space as they appear in a space-time frame, but a distinction between temporal and spatial relations." \cite{eddington}

Tim Maudlin \cite{maudlin0,maudlin1} argues along similar lines that the directed nature of time and the possibility to order all events along a temporal axis should be taken as a fundamental postulate in the scientific description of the world. He tries to reconstruct geometry from mathematical postulates based on such a linear ordering \cite{maudlin2}. Lee Smolin also subscribes to the idea that rather than removing time, it should be given greater emphasis in attempts to understand the physical world \cite{smolin}. Instead of relying on a sequential ordering of all events to achieve a universal definition of time, he argues that evolving laws of nature create the proper notion of time that is independent of external clocks.   

The present study goes along with the recent trend to rehabilitate time as a fundamental degree of freedom in proper physical models. Not only that, it is an attempt to restore its cognitive content. The aim is to clarify the conceptual status of time in the formalism, with the long term goal to resolve \emph{the problem of time}. The path towards that goal suggested here is not continued efforts to reconcile relativity and quantum theory \emph{as is}, but to dig deeper, reconsidering the fundamental notions of identity and change of objects, as well as the nature of physical states and their evolution.

The main difference between relativity and quantum theory is that the temporal and spatial variables are treated in the same way in relativity, but differently in quantum theory. In the latter theory, time $t$ is a precisely defined parameter, whereas position $x$ is an observable associated with an operator and subject to Heisenberg uncertainty. It is possible to call both $t$ and $x$ observables in relativity theory, since they can be associated with observer-dependent measurements. It is suggested here that this discrepancy between the two theories may be resolved in a broader theory containing a pair of observables $(x,t)$ supplemented with an evolution parameter $\sigma$.

This paper can be seen as a cognitive motivation for such a model. It is organized as follows. In section \ref{kant} the ideas to be developed are anchored in the Kantian tradition. These ideas are made more concrete in section \ref{correspondence}, in order to be useful as guidelines for building physical models. Two criteria for this purpose are introduced in section \ref{closure}, namely \emph{ontological completeness} and \emph{ontological minimalism}. Together they express a presumed one-to-one correspondence between forms of appearances and degrees of freedom in proper physical models. Ontological completeness demands that all subjective qualities of time should have formal counterparts in the model, which is crucial for this study. Epistemic physical states are introduced in section \ref{states}, and their temporal updates are described in section \ref{evolstates}. These updates are used to define the discrete \emph{sequential time} $n$ in section \ref{sequential}. To relate the universal $n$ properly with the observer-dependent relativistic space-time $(x,t)$, a distinction needs to be made between perceived objects and deduced \emph{quasiobjects}. Also, the role of knowledge of the past at some present time $n$ must be clarified. This is done in sections \ref{idquasi} and \ref{presentpast}. The necessity to introduce \emph{relational time} $t$ in addition to $n$ is discussed in section \ref{relationaltime}, together with some of the properties of $t$. Finally, the continuous evolution parameter $\sigma$ is defined in section \ref{evolp} from $n$. It is argued that $\sigma$ can play the role to evolve quantum mechanical states, but the details are left to an accompanying paper \cite{ostborn2}.

The philosophical background and the formalism it inspires is introduced in a manner that hopefully makes this paper self-contained, but at the same time as briefly as possible. More elaborate, but also more preliminary, discussion of the same material is found in Ref. \cite{ostborn}. The purpose of the philosophical part of the present discourse is to motivate a formalism that in turn can be used to motivate physical models. In other words, the philosophy is introduced a tool that may be useful to understand the foundations and structure of physics. Its merit is to be judged by its degree of usefulness for that purpose. Therefore, minimal emphasis is put here on a purely philosophical analysis, which would place the present ideas into historical context and compare them to other approaches in contemporary philosophy. Such a detailed analysis could be motivated if the present philosophical ansatz indeed turns out to be fruitful, in order to further sharpen the ideas, or to find inconsistencies.

\section{Kant's Copernican revolution applied to time}
\label{kant}

Physicists tend to make a clear distinction between time as it subjectively appears to us and time as a degree of freedom in their mathematical models. For example, in a debate on time with Bergson in 1922, Einstein stated \cite{bergson}:

\begin{quote}
\emph{There remains only a psychological time that differs from the physicist's.}
\end{quote}
Still, Einstein seems to have had a hard time coming to terms with this position. Rudolf Carnap recollected his attitude as follows \cite{carnap}:

\begin{quote}
\emph{Once Einstein said that the problem of the Now worried him seriously. He explained that the experience of the Now means something special for man, something essentially different from the past and the future, but that this important difference does not and cannot occur within physics. That this experience cannot be grasped by science seemed to him a matter of painful but inevitable resignation.}
\end{quote}

Rather than succumbing to resignation, an attempt is made in this paper to incorporate the cognitive structure that Einstein refers to into the physicist's time. In fact, this is not foreign to Einstein's own thinking in his youth, when he devised \emph{gedankenexperiment} based on cognition, relating the perceptions of observers in different states of motion and in free fall in order to motivate special and general relativity. The need to include the observer and her perceptions as a fundamental ingredient in physical models became even more pressing with the advent of quantum mechanics. Some physicists accepted this state of affairs and formulated the Copenhagen interpretation accordingly, whereas others did not - including Einstein.

The idea explored this paper is that this perspective should be accepted fully. The approach will not be limited to the use of the somewhat vague term 'observer' at the fundamental level of description, but aims to embrace the entire structure of subjective perceptions as a basis on which physical models are constructed. This attitude is similar to that of Kant, famously expressed in his \emph{Critique of Pure Reason} \cite{kant}:

\begin{quote}
\emph{Up to now it has been assumed that all our cognition must conform to the objects; but all attempts to find out something about them \emph{\emph{a priori}} through concepts that would extend our cognition have, on this presupposition, come to nothing. Hence let us once try whether we do not get farther with the problems of metaphysics by assuming that the objects must conform to our cognition.}
\end{quote}

The problems targeted in the present paper by such a Copernican revolution concern physics rather than metaphysics. In particular, the temporal form into which the cognized objects are placed is analyzed. Kant argues that such forms are given \emph{a priori}. The idea that the forms of appearances are given \emph{a priori} conforms to the idea that they can be used as a starting point to understand the form of physical law, which must also be seen as given \emph{a priori}. The following paraphrase of Kant may be used as a motto for the present study:

\begin{quote}
\emph{Hence let us once try whether we do not get farther with the problems of physics by assuming that its laws and the degrees of freedom they apply to must conform to the forms of appearances.}
\end{quote}

Kant viewed time as an essential form of appearance, stating that it is a necessary representation that grounds all intuitions.

\section{The correspondence between the world and the way it is perceived}
\label{correspondence}

\begin{figure}
\begin{center}
\includegraphics[width=80mm,clip=true]{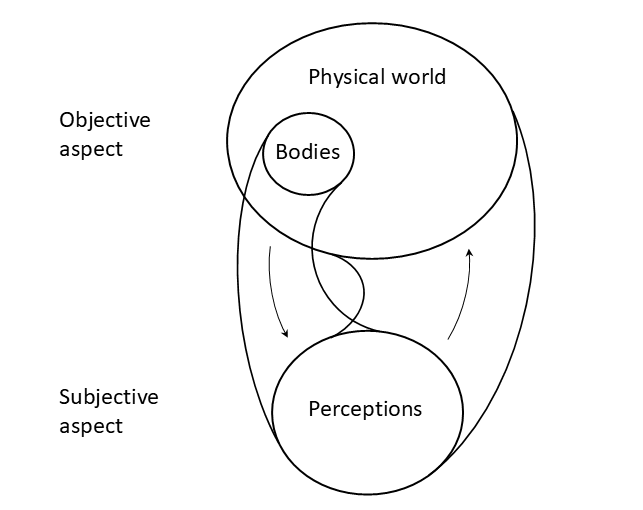}
\end{center}
\caption{Schematic illustration of the world view that inspires the assumptions used to motivate Born's rule. In this view, there is nothing objective in the physical world independent of the subjective, and the subjective is completely embedded in the objective in the sense that all perceptions can be correlated with physical states of bodies of perceiving subjects.}
\label{Figure0}
\end{figure}

From the viewpoint that leads to the formalization of time suggested in this paper, the subjective and objective aspects of the world are seen as equally fundamental, just like the relation between them, as illustrated in Fig. \ref{Figure0}. It is considered meaningless to talk about perception in itself without any objects to perceive, just like objects that cannot potentially be perceived by someone lack empirical meaning. The subjective and objective aspects of the world are thus seen as two sides of a coin: distinguishable but inseparable. This perspective is similar to Wheeler's idea of the participatory universe \cite{wheeler}.

In this setting, an \emph{object} is simply seen as an element of perception. It is a basic form of appearance that perceptions can be subjectively divided into a finite set of objects. The division is not necessarily spatial; we may hear a sound that we are able to divide into two tones with different pitch. Imagined things are also objects in this sense of the word, as well as abstract concepts.

\begin{figure}[tp]
\begin{center}
\includegraphics[width=80mm,clip=true]{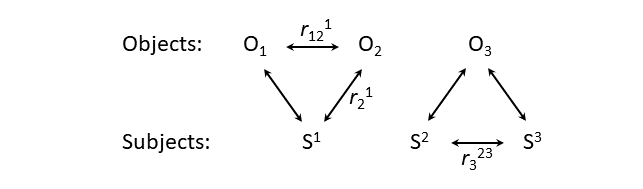}
\end{center}
\caption{The possibilities that one subject observes several objects and that several subjects observe one and the same object are postulated. The fact that subject $k$ knows about an object $O_{l}$ defines a relation $r_{l}^{k}$ between them. The circumstance that one subject knows two objects defines a relation $r_{ll'}^{k}$ between these objects. If two subjects know the same object, a relation $r_{l}^{kk'}$ between these subjects is defined.}
\label{Figure1}
\end{figure}

Just as any given subject can perceive several objects, any given object can be perceived by several subjects. The perception of two objects by the same subject may be taken as the definition of the circumstance that the two objects belong the same world. Similarly, the perception of the same object by two subjects may be seen as the definition of the fact that the two subjects live in the same world. A set of fundamental relationships is obtained between subjects and objects according to Fig. \ref{Figure1}.

To be able to identify objective physical law, and even to imagine and speak about it properly, we have to be able to interpret and generalize our subjective perceptions. A primary distinction between proper and improper such interpretations has to be assumed. This distinction transcends the perceptions themselves. Properly interpreted perceptions may be called \emph{knowledge}. The main idea at the heart of this paper is that the structure of knowledge and the structure of the world reflect each other. A thorough analysis of what can be known about the world and what cannot provides lessons about the world itself.

Physical law will be expressed in terms of states of knowledge and the changes they undergo. To do so, the ability to distinguish perceived objects belonging to the present and those belonging to the past has to be assumed. It must be possible to tell things happening now from memories. This is necessary to give meaning to the concepts of time and change in the present approach.

It should be stressed that even though the chosen perspective does not give any physical weight to transcendent \emph{things in themselves}, in the vocabulary of Kant, it does contain transcendent elements, which may be called \emph{objective}. Such elements are necessary in order to anchor the perceptions of a subject into something else, thus making it meaningful to talk about a world with certain properties and to study it scientifically. These transcendent elements include immutable forms of appearances, like time, proper interpretations of perceptions, or knowledge, and other perceiving subjects than oneself.

The idea that the structure of knowledge and the structure of the world reflect each other can be reformulated as the hypothesis that there is a one-to-one correspondence between the fundamental forms of appearances and the degrees of freedom in proper physical models of the world.

From this perspective, it becomes a worthwhile task to find the smallest set of fundamental forms of appearances to use in the physical model, and a corresponding smallest fundamental set of degrees of freedom. As a self-consistency check, each of these sets should make it possible to deduce the other. These ideas are made more concrete in the next section, using well-known physical examples.

\section{Examples of the correspondence}
\label{closure}

Consider Einstein's resignation, cited above, in the conclusion that the experience of \emph{the Now} cannot occur within physics. In the present vocabulary, this can be expressed as the fact that the temporal form of appearance \emph{now} has no immediate counterpart among the degrees of freedom of conventional physical models, and cannot be deduced from these models. According to the criteria of one-to-one correspondence and self-consistency, the set of degrees of freedom of the physical model has to be expanded.

The more radical view that time should be abandoned altogether as a fundamental degree of freedom in physics can be dismissed on the same grounds, given these criteria. Not only does this view make it impossible to account for any temporal forms of appearances at all, it leads to a de facto contradiction in terms, since the physicist who makes such a claim inevitably uses the category she denies: “Yesterday I realized that there is no need for time in physics, today I’m writing my result up, and tomorrow I’ll explain it at a seminar.”

Time is a premise in logic, which is a prerequisite for considering physical models at all. The prefix in the very words \emph{premise} and \emph{prerequisite} is temporal. The premise comes before the conclusion. The steps in our reasoning leading to the conclusion are like the ticking of a clock. The axioms used in a mathematical proof cannot be shuffled with the intermediate lines and the theorem, without destroying the logic. The order and direction of the elements in the proof are fundamental. Therefore, according to the chosen philosophical ansatz, directed time must be a fundamental degree of freedom in any proper physical model. 

To avoid this conclusion, it is possible to say that subjects and their ability to perceive, interpret and reason are not completely rooted in the physical world. We can view it from above, like angels, moving around using degrees of freedom we do not grant the world itself. The present approach avoids such mystical elements in the scientific world view.

\subsection{Ontological completeness}
\label{ocom}

The supposed need to incorporate time and the distinctions between past, present and future in a proper physical model can be seen as examples of the need to fulfill a principle of \emph{ontological completeness} (Fig. \ref{Figure2}). Namely, all subjectively perceivable distinctions correspond to distinctions in proper physical models. It is proposed that failure to fulfill this principle in a proposed physical model leads to inability to account for all experimental facts, or to wrong predictions.

Currently, informational approaches to quantum mechanics are popular, where information is understood in its binary form. The movement was initiated by Wheeler, with his catchphrase \emph{It from bit} \cite{wheeler2}, which morphed into \emph{It from qubit}. It is true that the simplest quantum mechanical systems have two possible states, and therefore may be called qubits. But it does not follow that all other quantum mechanical systems can be described as collections of qubits. Indeed, this is not the case, since there are fundamental systems which have more than two possible states, such as quarks with their three possible color charges. One might encode such states with a string of qubits, but it is conceptually wrong to use the code itself as a fundamental physical model. By definition, any code is accompanied by an external key. The distinction between systems with two, three or more possible states must in this case be built into the key, which becomes part of the physical model. The premise that this model is based solely on qubits breaks down.

Instead, according to ontological completeness, all kinds of variables that can be subjectively distinguished must be allowed into our physical models: binary, like the spin of the electron, discrete with more than two values, like color charges, directed, like time, undirected, like space, circular, like angles, and possibly continuous ones. None of these can be deduced from the others. They must therefore be considered equally fundamental.

\subsection{Ontological minimalism}
\label{omin}

\begin{figure}[tp]
\begin{center}
\includegraphics[width=80mm,clip=true]{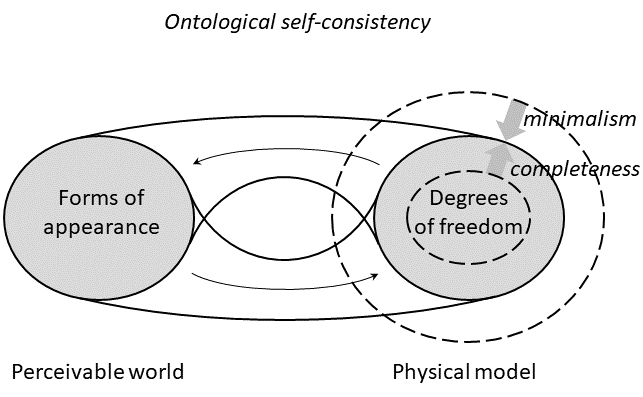}
\end{center}
\caption{It is assumed that a proper physical model should be complete, meaning that it contains all knowable degrees of freedom. The model should also be minimal, meaning that it does not contain any other degrees of freedom. All degrees of freedom of the model should follow from the forms of appearances, and it should be possible to derive all forms of appearances from the model. Such a model may be called ontologically self-consistent. It is argued that a model that is not ontologically self-consistent is insufficient or wrong. Such a loop of self-consisteny reflects the loop that describes the relation between the subjective and objective aspects of the world according to Fig. \ref{Figure0}.}
\label{Figure2}
\end{figure}

In order to achieve the assumed one-to-one correspondence between the fundamental forms of appearances and the degrees of freedom in proper physical models, a principle complementary to ontological completeness should be introduced, which may be called \emph{ontological minimalism} (Fig. \ref{Figure2}). It corresponds to the assumption that all distinctions in proper physical models correspond to subjectively knowable distinctions. Failure to exclude in a model distinctions that are not knowable is assumed to lead to wrong predictions.

It is a general lesson from the development of physics in the twentieth century that ontological dead weight in physical models should be thrown overboard. There turned out to be no knowable consequences of the introduction of the aether; therefore it was abandoned. More than that, an aether that defines an absolute spatial rest frame is inconsistent with the empirically confirmed predictions of special relativity. In a similar manner, wrong predictions are obtained when operations with no knowable effect are taken into account in physical models, such as the interchange of two identical particles in statistical mechanics, or the rotation of a spherically symmetric object. Physical law is ontologically picky.

This way of thinking led Heisenberg to the original mathematical formulation of quantum mechanics. The abstract of Heisenberg's breakthrough paper in 1925 simply reads \cite{heisenberg}, in English translation:

\begin{quote}
\emph{The present paper seeks to establish a basis for theoretical quantum mechanics founded exclusively upon relationships between quantities which in principle are observable.}
\end{quote}

Max Born also emphasized the significance of a similar principle. He used it to argue against the use of sharply defined continuous attribute values \cite{bornbohr}:

\begin{quote}
\emph{Modern physics has achieved its greatest successes by applying the methodological principle that concepts which refer to distinctions beyond possible experience have no physical meaning and ought to be eliminated. [...] I think that this principle should be applied also to the idea of physical continuity. Now consider, for instance, a statement like $x=\pi$ cm.; if $\pi_{n}$ is the approximation of $\pi$ by its first $n$ decimals, then the differences $\pi_{n}-\pi_{m}$ are, for sufficiently large $n$ and $m$, smaller than the accuracy of any possible measurement (even if it is conceded that this accuracy may be indefinitely improved in the course of time). Hence, statements of this kind should be eliminated.}
\end{quote}

The physical state introduced below takes this request to heart.

\section{Potential knowledge and physical states}
\label{states}

As noted above, from the chosen epistemic perspective an object is simply defined as an element of perception. Knowledge is assumed to consist of knowledge about a set of such objects and their attributes. An attribute can be defined as a set of qualities of an object that can be associated with each other and ordered. One quality in such a set can be called an attribute value. Red, green and blue are three different such attribute values that can be ordered into a spectrum that defines the attribute \emph{color}. Attributes may be internal or relational. An internal attribute, such as color, refers to the object itself, whereas a relational attribute, such as distance or angle, relates two or more objects. In some cases, it is possible to define pseudo-internal attributes from relational ones. One example is \emph{position}. This attribute refers to a single object, but is defined relationally via distances to other objects in a reference frame.

\begin{figure}[tp]
\begin{center}
\includegraphics[width=80mm,clip=true]{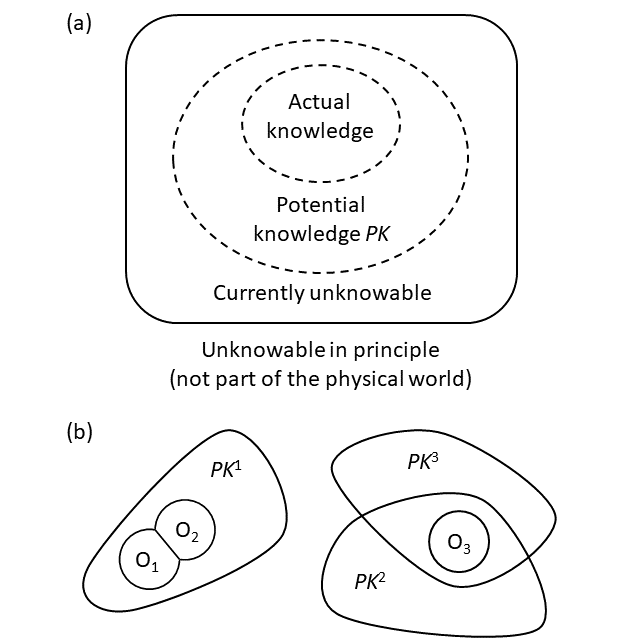}
\end{center}
\caption{(a) An onion of knowledge. The actual knowledge corresponds to the set of properly interpreted perceptions at some given time. The potential knowledge represents everything that is knowable in principle given the perceptions at that time. The solid outer boundary of the onion corresponds to the boundary of the physical world. There is such a boundary since the existence of proper and improper interpretation of perceptions is assumed. In other words, not everything conceivable is part of the world. (b) A given subject $S^{1}$ may have potential knowledge $PK^{1}$ about two objects $O_{1}$ and $O_{2}$, and two subjects $S^{2}$ and $S^{3}$ may have overlapping potential knowledge $PK^{2}$ and $PK^{3}$ about the same object $O_{3}$. Compare Fig. \ref{Figure1}. By definition $PK=\bigcup PK^{k}$.}
\label{Figure3}
\end{figure}

The considerations in this study rely on the possibility to use knowledge as a basis of a well-defined physical state. To fulfill such a role, the relevant knowledge must be precisely defined at any given time. However, the aware knowledge is quite fuzzy. Our attention shifts, we may have vague ideas, our interpretations of what we perceive may be correct or erroneous, or simply not settled. Therefore, a distinction needs to be made between the state of aware or \emph{actual knowledge}, which may be fuzzy, and the state of \emph{potential knowledge}, which is assumed to be well-defined. The potential knowledge at the given time is taken to correspond to all knowledge that may in principle be deduced from all subjective experiences at that time. It includes, for example, experiences that we do not become fully aware of until later. An onion of knowledge according to Fig. \ref{Figure3}(a) is obtained.

The existence of several perceiving subjects $S^{k}$ who may observe the same object $O_{l}$ has been assumed, as illustrated in Fig. \ref{Figure1}. One subject may have sharper eyes than the other, so that she has more knowledge about some attribute values of $O_{l}$. Different subjects may look at objects from different angles and also, of course, look at different objects. To avoid that an epistemic physical state depends on the perspective and perceiving power of different individuals and to enable that the state becomes a representation of the physical world as a whole, it clearly must correspond to the union

\begin{equation}
PK\equiv\bigcup_{k}PK^{k}
\label{cknow}
\end{equation}

of the individual potential knowledge $PK^{k}$ of each subject $S^{k}$ [Fig \ref{Figure3}(b)].

The outermost crust of the onion in Fig \ref{Figure3}(a) is the boundary between what is knowable in principle and what is not. The importance of this distinction in an epistemic approach to physics like the present one was emphasized by Eddington \cite{eddington}:

\begin{quote}
\emph{[W]e cannot limit the description to the immediate data of our own spasmodic obervations. The description should include nothing that is unobservable but a great deal that is actually unobserved.}
\end{quote}

\begin{quote}
\emph{The less stress we lay on the accident of parts of the world being known at the present era to particular minds, the more stress we must lay on the potentiality of being known to mind as a fundamental objective property of matter.}
\end{quote}

In the present treatment, it is assumed that potential knowledge $PK$ is always incomplete, meaning that at any given time there are objects or attribute values of objects that are unknown and currently unknowable, even though they may become known at some other time. This means that at each point in time it is impossible to know all objects that make up the world, or to know the precise value of each of their attributes. This idea conforms to the viewpoint of Born, cited above, that it is impossible in principle to know exactly the value of a continuous attribute, such as distance.

The reason for the assumed incompleteness of potential knowledge $PK$ is intuitively clear from Figure \ref{Figure0}. This knowledge is represented in the physical state of the bodies of all perceiving subjects, and the set of these bodies is considered to be a proper subset of the physical world. Thus, if $PK$ were complete, this knowledge would be represented in a proper subset of itself. This is not possible under the reasonable assumption that the states of the bodies and the state of the external world can vary independently. Similar hypotheses about the incompleteness of knowledge are reviewed by Szangolies under the poetic headline \emph{epistemic horizons} \cite{Szangolies}.

The supposed incompleteness of potential knowledge opens up a window to learn more about the world at the fundamental level as time goes by. Apart from perceiving new objects or learn their attribute values more precisely, this may correspond to the realization, using a figurative magnifying glass, that an observed object is in fact composite, that it consists of several parts.

It should be stressed that potential knowledge $PK$ at the conceptual level is considered to be independent of any particular representation or encoding of this knowledge, such as formal propositions or strings of bits. The possibility to distinguish knowledge from its representations follows from the assumption that subjective perceptions and insights about these perceptions are fundamental notions. At this conceptual level, it is possible to assign some set-theoretical relations between different states of knowledge according to Fig. \ref{Figure3} or Eq. [\ref{cknow}] from basic notions of having more or less knowledge, or sharing knowledge. However, to be able to perform more operations on states of knowledge, a particular representation of $PK$ called the physical state $S$ is introduced in the following.

\begin{figure}
\begin{center}
\includegraphics[width=80mm,clip=true]{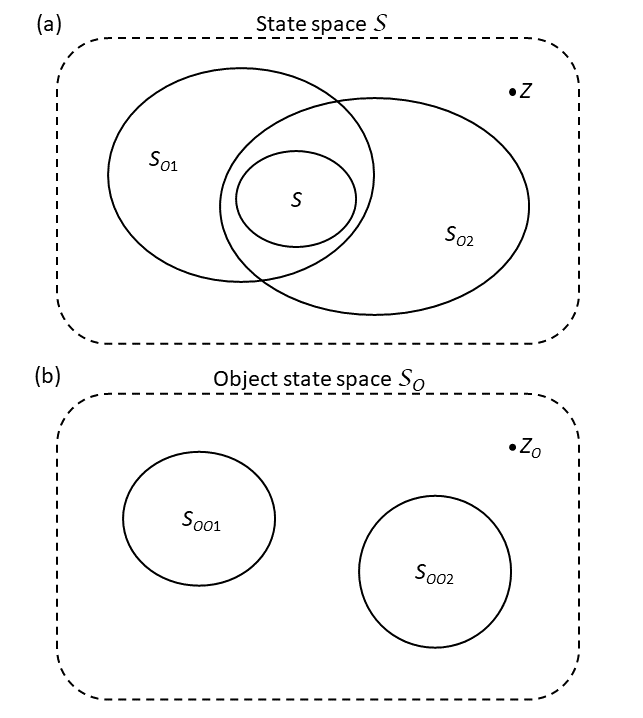}
\end{center}
\caption{(a) The physical state $S$ is the set of all exact states $Z$ of the world that are consistent with the incomplete potential knowledge $PK$. The object state $S_{O}$ is the set of all such $Z$ that are consistent with the potential knowledge specifically about object $O$. (b)  The object state can alternatively be represented as $S_{OO}$, the set of all exact states $Z_{O}$ of the object that are consistent with the potential knowledge about $O$. For two distinguishable objects $O_{1}$ and $O_{2}$ it is always true that $S_{O1}\cap S_{O2}\neq\varnothing$, whereas it always holds that $S_{OO1}\cap S_{OO2}=\varnothing$.}
\label{Figure4}
\end{figure}

Consider a hypothetical \emph{exact state} $Z$ where potential knowledge is completed, that is, all objects in the world are known, together with all their attributes and attribute values. Define a \emph{state space} according to $\mathcal{S}\equiv\{Z\}$. The \emph{physical state} $S$ that corresponds to potential knowledge $PK$ is then defined as $S\equiv \{Z_{PK}\}$, where $\{Z_{PK}\}$ is the set of exact states consistent with $PK$, or the set of such states that cannot be excluded given $PK$. It was argued that potential knowledge is always incomplete. Therefore $S$ will always contain several elements $Z$, according to Fig. \ref{Figure4}(a).

It is a curious consequence of this definition of the physical state $S$ that its boundary $\partial S$ in general cannot be precisely known. That would entail a precise discrimination between pairs of exact states $Z$ and $Z'$ on either side of $\partial S$, but these elements of $\mathcal{S}$ are no actual physical states at all, and cannot be known and compared, given the incomplete knowledge that defines $S$ in the first place. 

Most often, physics deals with the study of specific objects, rather than with the universe as a whole. Any proper subset of the world can be regarded to be such an object. The \emph{object state} $S_{O}$ is defined as the physical state that would result if the knowledge about all the other objects were erased. More precisely, $S_{O}$ is the union of all exact states $Z\in\mathcal{S}$ that do not contradict the fact that $O$ exists, or the potential knowledge of its internal attributes. It holds that $S\subset S_{O}$, according to Fig. \ref{Figure4}(a).

In the following, it will be convenient to embed the object state in another state space $\mathcal{S}_{O}\equiv\{Z_{O}\}$, the \emph{object state space}, where the \emph{exact object state} $Z_{O}$ corresponds to a completion of the potential knowledge about a specific object $O$. This potential knowledge is taken to include internal and pseudo-internal attributes of $O$. Pseudo-internal attributes like position are included in order to be able to distinguish two seemingly identical objects at different spatio-temporal locations. The corresponding object state is denoted $S_{OO}$, and $S_{OO}\subset\mathcal{S}_{O}$ according to Fig. \ref{Figure4}(b).

For the states $S_{OO1}$ and $S_{OO2}$ of two objects $O_{1}$ and $O_{2}$ that are known to be different at a given time it holds that $S_{OO1}\cap S_{OO2}=\varnothing$. If this relation does not hold, on the other hand, it has no epistemic meaning to say that there are two objects, since they cannot be distinguished, neither by their internal attributes, nor by their relational attributes like the spatial position. The situation is different when the object states are embedded in the full state space $\mathcal{S}$. Then it is always true that $S_{O1}\cap S_{O2}\neq\varnothing$, since $S\subset S_{O1}$ and $S\subset S_{O2}$, according to Fig. \ref{Figure4}(a). This representation of object states is clearly less intuitive.

\section{Evolution of physical states}
\label{evolstates}

Suppose that the perception of some subject $k$ changes, so that $PK^{k}\rightarrow \tilde{PK}^{k}$. Such a change corresponds to a binary distinction between before and after, and thus makes it possible to assign a discrete time $n$ to the potential knowledge $PK^{k}(n)$ before, and a time $n+1$ to the potential knowledge $PK^{k}(n+1)\equiv\tilde{PK}^{k}$ after the change.

Assume an operational materialistic world view in the sense that any perceived change corresponds to a knowable change in the state of the body of the perceiving subject. This amounts to the assumption that all subjective perceptions have a neural correlate. Then the perceived change $PK^{k}(n)\rightarrow PK^{k}(n+1)$ must correspond to a change in the state of the world of which the body of $k$ is a part. Therefore the collective potential knowledge $PK$ must also change, so that

\begin{equation}
PK(n)\neq PK(n+1).
\label{newdiff}
\end{equation}

It can be argued that all perceived changes can be put into this binary form, even though some processes are continuous, like a ball following a ballistic trajectory. In such cases there is nevertheless a series of moments at which the observer realizes that the position of the ball has changed. These moments define the discrete temporal updates. This matter will be further discussed in sections \ref{sequential} and \ref{presentpast}.

\subsection{The evolution operator}

\begin{figure}[tp]
\begin{center}
\includegraphics[width=80mm,clip=true]{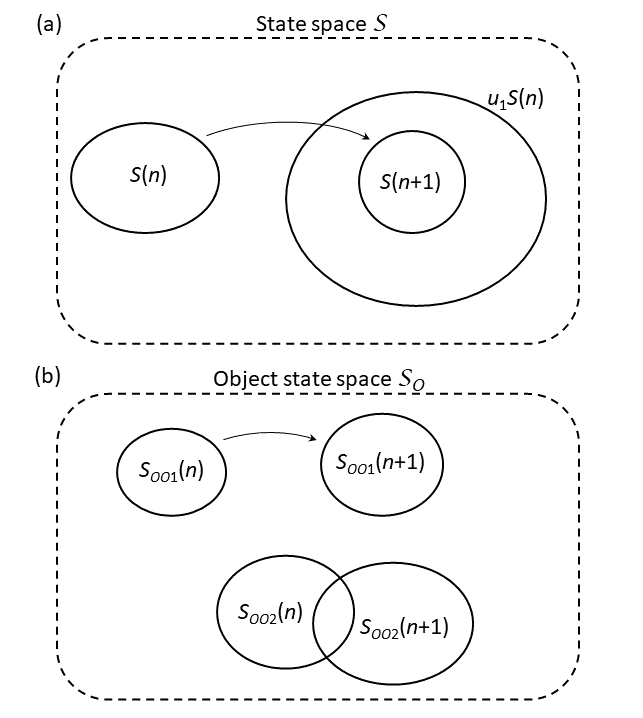}
\end{center}
\caption{(a) The evolution of the physical state $S(n)$ is such that $u_{1}S(n)$ and $S(n+1)$ do not overlap $S(n)$. (b) This may or may not be true when it comes to the evolution of object states $S_{OO}(n)$.}
\label{Figure5}
\end{figure}

Any change of $PK$ makes it possible to distinguish the physical state before and after, so that

\begin{equation}
S(n)\cap S(n+1)=\varnothing,
\label{snew}
\end{equation}
as illustrated in Fig. \ref{Figure5}(a). It is possible to write

\begin{equation}
S(n+1)\subseteq u_{1}S(n),
\label{sevolution}
\end{equation}
where the evolution operator $u_{1}$ is defined by the condition that $u_{1}S(n)\subseteq\mathcal{S}$ is the smallest set for which Eq. (\ref{sevolution}) is guaranteed to be fulfilled. Physical law can thus be expressed in part as a mapping $u_{1}:\mathcal{P}(\mathcal{S})\rightarrow \mathcal{P}(\mathcal{S})$ from the power set $\mathcal{P}(\mathcal{S})$ of state space $\mathcal{S}$ to itself.

Since $S(n+1)\subseteq u_{1}S(n)$ cannot overlap $S(n)$ according to Eq. (\ref{snew}), the evolution operator $u_{1}$ should be defined such that $u_{1}S(n)$ and $S(n)$ do not overlap either:

\begin{equation}
S(n)\cap u_{1}S(n)=\varnothing.
\label{evsnew}
\end{equation}

\begin{figure}[tp]
\begin{center}
\includegraphics[width=80mm,clip=true]{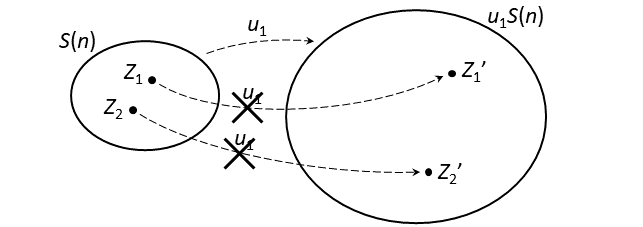}
\end{center}
\caption{The evolution operator $u_{1}$ can only be applied to a physical state $S$, which means that it cannot be applied to its individual elements $Z$. The same goes for the evolution operator $u_{O1}$, which applies to object states $S_{OO}$, but not to its elements $Z_{O}$. Such a 'holistic' evolution is qualitatively different from the evolution of an ensemble in classical phase space, which is defined by the evolution of each of its elements. These correspond to possible physical states, in contrast to the exact states $Z$.}
\label{Figure6}
\end{figure}

The principle of ontological minimalism discussed in section \ref{omin} implies that the evolution cannot be reduced to an element-wise mapping $u_{1}:\mathcal{S}\rightarrow \mathcal{S}$. That is, the exact states $Z$ are not in the domain of $u_{1}$, meaning that the expression $u_{1}S=u_{1}\{Z_{i}\}=\{u_{1}Z_{i}\}$ is improper, as illustrated in Fig. \ref{Figure6}. The reason is that the 'states' $Z_{i}$ are unphysical if they are regarded individually, since it is argued that knowledge is always incomplete. According to the principle of ontological minimalism, physical law cannot be properly described by referring to entities or distinctions that are unknowable in principle, such as the elements $Z$ of $\mathcal{S}$.

In order to give epistemic meaning to the statement that there are two possible values $a_{1}$ and $a_{2}$ of an attribute to an object $O$, but none of them can be excluded at time $n$ because of the incompleteness of potential knowledge, it must be possible in principle to learn later which of these values actually applies to $O$, and both options should be on the table. If this new knowledge were to be dictated by physical law, the physical state $S(n)$ would have to contain exact states $Z$ were the value is either $a_{1}$ and $a_{2}$, but $u_{1}S(n)$ should only contain exact states where the value is $a_{1}$, say. However, in that case both options would no longer be on the table; it would be certain at time $n$ that value $a_{1}$ will be found, contrary to assumption. It is concluded that \emph{state reductions}

\begin{equation}
u_{1}S(n)\rightarrow S(n+1)\subset u_{1}S(n)
\label{sred}
\end{equation}
must be allowed, where the state $S(n+1)$ cannot be predicted from $S(n)$ or $u_{1}S(n)$. This is illustrated in Fig. \ref{Figure5}(a). Even though deterministic evolution $S(n+1)=u_{1}S(n)$ should also be allowed, physical law as a whole becomes indeterministic. The state reduction in Eq. (\ref{sred}) can be seen as a generalization of a wave function collapse in quantum mechanics.

Two more general qualities of the evolution of physical states can be motivated by similar plausibility arguments. First, the evolution should not be \emph{different} depending on the \emph{amount} of potential knowledge available in a given situation. That is,

\begin{equation}
S'(n)\subset S(n)\Rightarrow u_{1}S'(n)\subset u_{1}S(n).
\label{samount}
\end{equation}
This hypothesis expresses an elementary form of reductionism. The behavior of a system can be understood by a closer study, by resolving and determining the behavior of its parts. In a similar vein, it is reasonable to assume that if $S$ and $S'$ overlap, then $u_{1}S$ and $u_{1}S'$ also overlap:

\begin{equation}
S\cap S'\neq\varnothing\Rightarrow u_{1}S\cap u_{1}S'\neq\varnothing.
\label{soverlap}
\end{equation} 
This can be regarded as an assumption that the evolution $u_{1}$ is a function in a subjective sense: two states that are subjectively the same cannot evolve into states that are subjectively different. This rule seems necessary given the argument above that it is impossible to know the boundary $\partial S$ of the physical state precisely. If different such boundaries could lead to divergence among the evolved states, then it would be impossible to make predictions, or to determine physical law at all.

\subsection{The object evolution operator}

It is possible to define an evolution operator $u_{O1}$ that applies to object states $S_{OO}$ in an analogous way as the universal evolution operator $u_{1}$ is defined in relation to Eq. (\ref{sevolution}). Namely,

\begin{equation}
S_{OO}(n+1)\subseteq u_{O1}[S(n)]S_{OO}(n),
\label{soevolution}
\end{equation}
where $u_{O1}[S(n)]S_{OO}(n)\subseteq\mathcal{S}_{O}$ is the smallest set for which Eq. (\ref{soevolution}) is always fulfilled. The object evolution operator $u_{O1}$ is a mapping $u_{O1}:\mathcal{P}(\mathcal{S}_{O})\rightarrow \mathcal{P}(\mathcal{S}_{O})$ from the power set $\mathcal{P}(\mathcal{S}_{O})$ of object state space $\mathcal{S}_{O}$ to itself.

The operator $u_{O1}$ depends on the state $S(n)$ of the entire world, since any object $O$ is always related to the world to which it belongs. The overwhelming success of the reductionistic approach in science makes it possible to assume that $u_{O1}$ can always be expressed in terms of its action on a set of elementary particles. Then it attains a general form which is independent of time $n$ and the particular object state $S_{OO}$ on which it acts.

Several qualities of the object evolution operator $u_{O1}$ can be taken to be the same as those of the universal evolution operator $u_{1}$. For the same reason as in the universal case, object state reductions must be allowed:

\begin{equation}
u_{O1}S_{OO}(n)\rightarrow S_{OO}(n+1)\subset u_{O1}S_{OO}(n).
\label{sored}
\end{equation}

Also,

\begin{equation}
S_{OO}'(n)\subset S_{OO}(n)\Rightarrow u_{O1}S_{OO}'(n)\subset u_{O1}S_{OO}(n).
\label{sooamount}
\end{equation}
and

\begin{equation}
S_{OO}\cap S_{OO}'\neq\varnothing\Rightarrow u_{O1}S_{OO}\cap u_{O1}S_{OO}'\neq\varnothing.
\label{sooverlap}
\end{equation}

There is a crucial difference between $u_{O1}$ and $u_{1}$, though, as expressed in Fig. \ref{Figure5}. The evolved object states $u_{O1}S_{OO}(n)$ and $S_{OO}(n+1)$ may or may not overlap $S_{OO}(n)$, whereas $u_{1}S(n)$ never overlaps $S(n)$.

\subsection{The pointer and dial of a clock}
There must be at least one object $O_{1}$ such that $S_{OO1}(n)$ and $S_{OO1}(n+1)$ do not overlap. The perceivable change of such an object defines the temporal update $n\rightarrow n+1$. There must also be at least one object $O_{2}$ such that $S_{OO2}(n)$ and $S_{OO2}(n+1)$ overlap. Such objects preserve the identity of the world in which the change takes place. These statements may be motivated as follows.

\begin{figure}[tp]
\begin{center}
\includegraphics[width=80mm,clip=true]{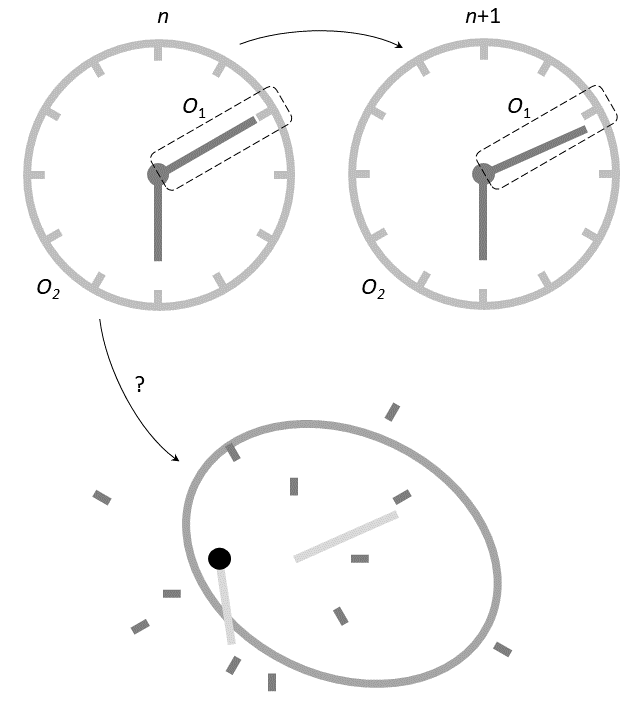}
\end{center}
\caption{A clock must have two components: a pointer that changes and a dial that stays the same. These components correspond to objects $O_{1}$ and $O_{2}$ in Fig. \ref{Figure5}(b), respectively. If there is no part of the clock that stays the same when the pointer moves, it is impossible to say that it is the same clock before and after this movement, and its function as a clock is lost. More generally, there must always be an object $O_{1}$ that subjectively changes in order to define a temporal update $n\rightarrow n+1$, and there must always be another object $O_{2}$ that does not, in order to speak about the passage of time in a given world.}
\label{Figure7}
\end{figure}

When something changes something else must stay the same in order to make it epistemically meaningful to relate what comes after with what was before. A leaf may fall from the tree, but the stem of the tree does not move. When the leaf falls, the immobility of the stem helps preserving the identity of the tree, and of the entire world. When the leaf has fallen, the tree may be cut down, and the leaf resting on the ground helps preserving the identity of the world. Even if everything around us seem to change at once, some internal objects may stay the same, such as our mood and memories. In this way, the evolution of a given, identifiable world may be compared to walking: when one foot is lifted, the other is resting on the ground. These considerations are implicit in the definition of a clock, as expressed in Fig. \ref{Figure7}.

\section{Sequential time}
\label{sequential}

The temporal update $n\rightarrow n+1$ is associated with a change of perception by some subject $k$. To define a longer sequence $\{n,n+1,n+2,\ldots\}$ of temporal updates, the interplay of changes of perceptions of different subjects must be considered. It will be argued that such a sequence, with an ordering of events that everyone can agree upon, may exist despite the relativity of simultaneity. A directed sequence of that type is proposed as a basis for a universal \emph{sequential time}.

It was noted in section \ref{intro} that some researchers question the notion of a universal time because it would require a clock external to the universe, which would be a contradiction in terms. The present construction circumvents that objection since it does not rely on the measure of time provided by a clock, but only on the ordering of perceived events. These perceptions are considered fundamental to the physical world, as expressed in Figs. \ref{Figure0} and \ref{Figure2}, together with the associated forms of appearances, such as the flow of time. Even so, these perceptions are assumed to correlate with evolving states of the body associated with a perceiving subject, allowing for an operational form of materialism. It should be noted, though, that these bodies are themselves considered to be nothing more or less than objects of potential perception.   

A change potentially perceived by subject $k$ corresponds to an event $e^{k}$:

\begin{equation}
\begin{array}{ll}
e^{k}: & PK^{k}\rightarrow \tilde{PK}^{k}.
\end{array}
\label{event}
\end{equation}

Such an event implies an update $PK(n)\rightarrow PK(n+1)\neq PK(n)$ of the collective potential knowledge $PK$, according to Eq. (\ref{newdiff}). Given the assumption that the directed flow of time is a fundamental form of appearance that should be reflected in physical law, it is possible to order the set of events $\{e_{j}^{k}\}_{j}$ potentially perceived by $k$ in a sequence 

\begin{equation}
\{\ldots, e_{j}^{k},e_{j+1}^{k},e_{j+2}^{k},\ldots\}.
\end{equation}

The basic question is whether the set $\{e_{j}^{k}\}_{jk}$ of such events perceived by the set of all subjects 1) necessarily forms a temporally well-defined collective sequence

\begin{equation}
\{\ldots,e_{j}^{k},e_{j'}^{k'},e_{j+1}^{k},e_{j''}^{k''},e_{j'+1}^{k'},\ldots\},
\label{sequence}
\end{equation}
or whether 2) it may be arranged in such a sequence, or whether 3) such a collective sequential arrangement is impossible. In the first case, a universal sequential time is automatically defined, in the second case it may be defined, and in the third case it cannot be defined.

The events in the set $\{e_{j}^{k}\}_{jk}$ would necessarily form a temporal sequence if any two subjects $k$ and $k'$ were able to determine the temporal order of any pair of events $e_{j}^{k}$ and $e_{j'}^{k'}$ by means of communication. This would mean that subject $k$ perceives $e_{j}^{k}$ and sends a signal to $k'$ announcing this fact. If subject $k'$ receives this signal before she perceives event $e_{j'}^{k'}$, it is clear that $e_{j}^{k}$ should be placed before $e_{j'}^{k'}$ in the temporal sequence (\ref{sequence}). However, such communication is limited by the speed of light. In other words, such ordering of events is possible if they are temporally separated.

In such a spatio-temporal representation each event $e^{k}$ is appropriately placed along the world line of the body of the subject $k$ who perceives it, even though the proper \emph{interpretation} of her perception may be that it is associated to a distant event, such as a supernova explosion thousands of light years away. This perspective is consistent with the epistemic approach adopted in the present paper, in the spirit of Kant's Copernican revolution. These matters are further discussed in section \ref{contded}.

When the events $e_{j}^{k}$ and $e_{j'}^{k'}$ are spatially separated, on the other hand, it is impossible to judge their temporal ordering by means of signalling. One might want to argue that since their ordering cannot be empirically determined, they should be regarded as simultaneous. Allowing simultaneous events in a sequence like that in Eq. (\ref{sequence}) is unproblematic from a conceptual point of view. It would still be possible to use such a sequence to define universal time.

However, it is inconsistent to assign simultaneity to all spatially separated events. Consider three events $e_{1}$, $e_{2}$ and $e_{3}$, forming a triangle when represented in space-time. If $e_{1}$ is perceived by a subject very far from the subjects that perceive $e_{2}$ and $e_{3}$, it may happen that $e_{1}$ is spatially separated from $e_{2}$ and $e_{3}$, but these latter two events are temporally separated. Then it is cleary impossible to say that $e_{1}$ is simultaneous with $e_{2}$ and $e_{3}$, since then $e_{2}$ and $e_{3}$ would also be simultaneous, contrary to assumption.

\begin{figure}[tp]
\begin{center}
\includegraphics[width=80mm,clip=true]{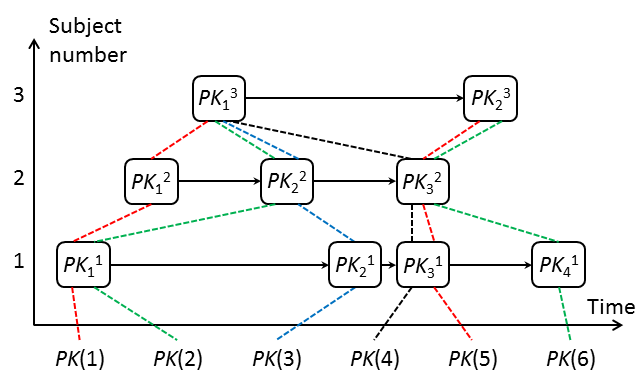}
\end{center}
\caption{The potential knowledge $PK(n)$ is updated each time any of the individual states of knowledge $PK_{j}^{k}$ are updated. For example, $PK(2)$ is the union of $PK_{1}^{1}$, $PK_{2}^{2}$, and $PK_{1}^{3}$. This state is updated to $PK(3)$ when $PK_{1}^{1}\rightarrow PK_{2}^{1}$. If two individual updates have space-like separation, they can sometimes be considered to occur simultaneously, as the updates $PK_{2}^{1}\rightarrow PK_{3}^{1}$ and $PK_{2}^{2}\rightarrow PK_{3}^{2}$. Overlap between individual states of potential knowledge typically occur according to Fig. \ref{Figure3}(b), but this is not shown here for clarity. }
\label{Figure8}
\end{figure} 

Therefore, well-defined temporal ordering of spatially separated events must be allowed. It is postulated that such an ordering exists, still allowing for some spatially separated events to be simultaneous. The set of all individually perceived events $\{e_{j}^{k}\}_{jk}$ may then be used as a collection of time stamps that define a sequence

\begin{equation}
\{PK(n), PK(n+1), PK(n+2) \ldots\}
\label{pksequence}
\end{equation}
of distinguishable collective states of potential knowledge. The suggested relation between the updates of individual and collective states of potential knowledge is illustrated in Fig. \ref{Figure8}. The sequence (\ref{pksequence}) in turn defines the sequence

\begin{equation}
\{S(n), S(n+1), S(n+2) \ldots\}
\label{ssequence}
\end{equation}
of distinct physical states $S(n)\cap S(n+1)=\varnothing$ according to Eq. (\ref{snew}), where $n$ defines \emph{sequential time}. 

At this stage it is evident that a universal sequential time $n$ does not exist by necessity, since the existence of the ordering of the events that defines it sometimes has to be postulated. It may instead be argued that such a sequential time is impossible, referring to the ambiguity of the ordering of spatially separated events that follows from special relativity. The apparent conflict between the postulated sequential time $n$ and the relativity of simultaneity will be addressed in section \ref{revsim}, where it is claimed that relativity does not apply to $n$, but to another aspect of time, called relational time $t$.

\section{Identifiable objects and quasiobjects}
\label{idquasi}

The postulated existence of sequential time provides some underlying structure to our perceptions. This structure becomes objective in the sense that it is collective, transcending the perceptions of the individual subject. However, the picture of the physical world developed so far may still be dismissed as a kaleidoscopic sequence of perceived images, lacking the inner structure or continuity that is evident from experience. It is clear from experience that physical laws can be deduced that act on objects that preserve their identity as time passes. These objects can often be modelled as 'being there' regardless whether we observe them or not. In this section, the concepts of \emph{identifiable objects} and \emph{quasiobjects} are introduced in order to address these issues.

It was argued in relation to Fig. \ref{Figure1} that the assumption that two subjects $k$ and $k'$ can perceive \emph{the same} object $O$ can be seen as an expression of the circumstance that they live in the same world. According to the definition of the object state $S_{OO}$ provided in relation to Fig. \ref{Figure4}(b), it holds that $S_{OO}^{k}(n)$ and $S_{OO}^{k'}(n)$ overlap if $O$ is indeed simultaneously observed by $k$ and $k'$ at some sequential time $n$. Conversely, if $S_{OO}^{k}(n)$ and $S_{O'O'}^{k'}(n)$ overlap, then $O$ and $O'$ must be identified as the same object, since they cannot be subjectively distinguished.

\begin{figure}[tp]
\begin{center}
\includegraphics[width=80mm,clip=true]{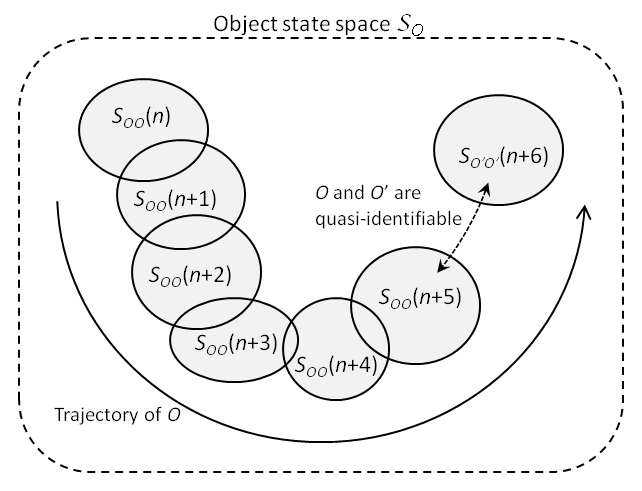}
\end{center}
\caption{The evolution of the state $S_{OO}$ of an object $O$ that is identifiable in the time interval $[n,n+5]$. The object $O'$ at time $n+6$ cannot be identified with object $O$ at time $n+5$, and is therefore given a separate name. If $O$ can nevertheless be modeled as being composed of identifiable quasiobjects which may evolve so that $S_{OO}(n+6)=S_{O'O'}(n+6)$, then the identification $O=O'$ can be made, and it is possible to say that $O$ is quasi-identifiable in the time interval $[n,n+6]$.}
\label{Figure9}
\end{figure}

In an analogous way, an object $O$ will be called \emph{identifiable} at time $n$ if and only if $S_{OO}(n)$ and $S_{OO}(n+1)$ overlap, that is

\begin{equation}
S_{OO}(n)\cap S_{OO}(n+1)\neq \varnothing.
\label{id}
\end{equation}
If object $O$ is identifiable at all times $n, n+1,\ldots,n+m$, then it is said to be identifiable in the time interval $[n,n+m]$. This condition gives meaning to the statement that \emph{the same} object is tracked from time $n$ to time $n+m$ (Fig. \ref{Figure9}).

The object $O_{1}$ that defines the temporal update $n\rightarrow n+1$ according to Fig. \ref{Figure7} is not identifiable at time $n$. It is nevertheless possible to say that it preserves its identity after the change associated with this update. This is so since it is assumed that it is possible to \emph{model} any object as being composed of, or corresponding to, a set of identifiable \emph{quasiobjects}. A quasiobject $QO$ is an object that is not directly perceived, but is used to express in an efficient and general way the physical law that governs the evolution of the objects that are actually perceived. Such an abstract object is identifiable if it can be \emph{modeled} as being identifiable in a proper expression of physical law.

According to the discussion in relation to Fig. \ref{Figure5}, an event $e^{k}$, as defined in Eq. (\ref{event}), is always associated with a potentially perceived change of the state of some object $O_{1}$. If this object can be modelled as being composed of identifiable quasiobjects, it may be called \emph{quasiidentifiable} and the event can be characterized as follows:

\begin{equation}
\begin{array}{lll}
e^{k}: & S_{OO1}(n)\rightarrow S_{OO1}(n+1), & S_{OO1}(n)\cap S{OO1}(n+1)=\varnothing.
\end{array}
\label{event2}
\end{equation}
Such an observed event can be interpreted as being caused by an event that changes the state of a quasiobject that is part of $O_{1}$. For example, the click by a Geiger counter can be interpreted as corresponding to the event that radiation enters the tube and ionizes an atom of the gas that fills it. This event is not perceived but deduced, and may thus be called a \emph{quasievent}:

\begin{equation}
\begin{array}{lll}
qe: & S_{QOQO1}(n)\rightarrow S_{QOQO1}(n+1), & S_{QOQO1}(n)\cap S_{QOQO1}(n+1)=\varnothing,
\end{array}
\label{qevent2}
\end{equation}
where $QO_{1}$ is the changing quasiobject that is interpreted as causing the change of the object $O_{1}$ that defines the observed event (\ref{event2}).

The notion of quasiobjects is unavoidable in the present world view, because of the possibility to learn more about perceived objects due to the supposed incompleteness of knowledge. This means that that some of these objects should be imagined to be composed of smaller parts. These parts are quasiobjects before they are actually seen.

Some identifiable quasiobjects can never be seen, like elementary particles, even though they can sometimes be put in one-to-one correspondence to a perceived object, such as a trace in a bubble chamber or some other detector. A large object like the sun can be perceived, of course, but may become an identifiable quasiobject during a limited period of time, for instance if there is nobody on the other side of the world who sees the sun after sunset. It is still possible to say that it is the same sun that rises the next morning since it can be modeled as a collection of identifiable elementary particles in a way that conforms with all actual observations of the sun.

Even though quasiobjects are indispensable in order to create a model of the observed world, and to formulate physical law in a simple and general way, it is important to note that from the strict epistemic perspective adopted here, they do not add any new knowledge about the physical state, which corresponds to the properly interpreted potential perceptions at a given sequential time $n$. All that can be known about the quasiobjects is deduced from this potential knowledge. This knowledge is primary and the quasiobjects are secondary abstractions. This perspective is opposite to the conventional viewpoint in physics where the quasiobjects are seen as primary - they are really out there - and the perceptions that give rise to knowledge are secondary, capturing only a small subset of the information stored in the state of all the quasiobjects.

\section{Present and past}
\label{presentpast}

From the cognitive perspective adopted in this paper, the past exists as memories and records that are potentially perceived in the present. Perceptionwise, there is no legitimate viewpoint from which the sequence of potential states of knowledge ($\ref{pksequence}$) or physical states (\ref{ssequence}) can be accessed, since no subject can reach such an extra-temporal vantage point. Therefore, the past somehow ought to be represented as part of the present. This is the subject of the present section. Adhering to the principle of ontological completeness discussed in section \ref{ocom}, such a representation must be introduced in proper physical models.

Consider the state $S_{OO}(n)$ of some object $O$ at the present time $n$. There may also be a memory at time $n$ of its preceding state $S_{OO}(n-1)$. The object state $S_{OO}(n-1)$ in itself cannot be considered to be a part of the physical state $S(n)$, simply because the two states refer to different sequential times. Rather, the memory of object $O$ at time $n$ is taken to correspond to another object $O'$, the knowledge of which is part of $S(n)$. The state at time $n$ of the remembered object may be written $S_{O'O'}(n)=M(S_{OO}(n-1))$, where $M(\ldots)$ denotes the potential memory of the state within brackets.

\begin{figure}[tp]
\begin{center}
\includegraphics[width=80mm,clip=true]{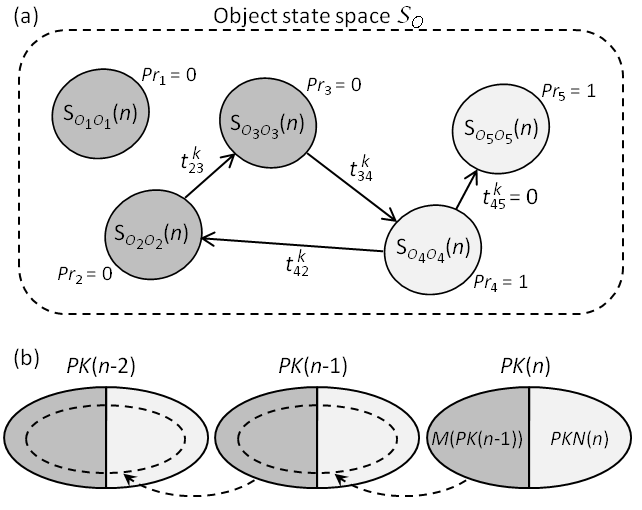}
\end{center}
\caption{The roles of sequential time $n$ and relational time $t$. (a) At any given time $n$ the state of potential knowledge $PK(n)$ that corresponds to the physical state $S(n)$ contains several objects $O_{l}$ with physical object states $S_{O_{l}O_{l}}(n)$. Some of these are perceptions of the present (light gray), and some are memories of the past (dark gray). The temporal distances $t_{ll'}^{k}$ measured by subject $k$ relate any two such objects $O_{l}$ and $O_{l'}$ and the knowledge of their values is part of $PK^{k}(n)$ and thus of $PK(n)$. (b) Sequential time labels different states of potential knowledge $PK(n)$. The set of memories $M(PK(n-1))\subseteq PK(n)$ at time $n$ consists of all dark gray objects $O_{l}$ in panel (a). This set refers back to the preceding state $PK(n-1)$. Memories may be imperfect, which is illustrated by the fact that the dashed ovals are proper subsets of $PK(n-1)$ and $PK(n-2)$, respectively. The backward references define a unique directed ordering of instants $n$, reflecting the perceived flow of time.}
\label{Figure10}
\end{figure}

Some objects perceived at time $n$ are clearly present objects and some are memories of past objects. It is therefore possible to define a binary \emph{presentness attribute} $Pr$ that applies to all objects, such that $Pr=1$ for present objects and $Pr=0$ for memories of past objects, as illustrated in Fig. \ref{Figure10}(a). The value of $Pr$ defines the object state $S_{OO}$ together with the values of an array of other attributes, such as charge and position. As noted above, attributes may be \emph{internal} or \emph{relational}. Presentness is clearly an internal attribute that refers to the object $O$ itself, just like charge, whereas position is relational.

The assumed incompleteness of knowledge implies that the knowledge about the value of any attribute $A$ may be incomplete, meaning that there are more than one value that is not excluded by the potential knowledge about the object. This goes for the presentness attribute $Pr$ as well. Imagine, for example, that you look at a tree and contemplate the leaves rustling in the wind. You know that the leaves move along given trajectories, but you cannot tell which of their positions belong to the present and which belong to the immediate past. The experience is temporally holistic, in a sense.

From this perspective, the potential knowledge $PK(n)$ that corresponds to the physical state $S(n)$ may be divided into two parts

\begin{equation}
PK(n)=PKN(n)\cup M(PK(n-1)),
\label{pkparts}
\end{equation}
where $PKN(n)$ corresponds to the knowledge at time $n$ about objects with $Pr=1$, and $M(PK(n-1))$ corresponds to the knowledge at time $n$ about objects with $Pr=0$ [Fig. \ref{Figure10}(b)]. One might want to add a third part of $PK(n)$ in Eq. (\ref{pkparts}), corresponding to the potential knowledge about those objects for which the value of $Pr$ is uncertain. However to make the notation as simple as possible, such objects are assumed to belong to both $PKN(n)$ and $M(PK(n-1))$, so that the knowledge about such temporally ambiguous objects corresponds to $PKN(n)\cap M(PK(n-1))$.

The potential memories $M(PK(n-1))$ of the preceding sequential time $n-1$ may be perfect or imperfect, so that

\begin{equation}
M(PK(n-1))\subseteq PK(n-1)
\label{sloppymemory}
\end{equation}
if the changes in the presentness attribute $Pr$ of some objects when going from $PK(n-1)$ to $M(PK(n-1))$ are disregarded. The knowledge denoted by $M(PK(n-1))$ corresponds by definition to \emph{proper} memories, so that $M(PK(n-1))$ always refers to $PK(n-1)$ in a clear-cut way. This reference can be illustrated as an arrow pointing from  $M(PK(n-1))$ to $PK(n-1)$, as shown in Fig. \ref{Figure10}(b). In the same way, there is an arrow pointing from $M(PK(n-2))$ to $PK(n-2)$. In this way a set of arrows is defined that creates a unique directed ordering of the elements in the set $\{PK(n)\}_{n}$ according to

\begin{equation}
\ldots\leftarrow PK(n-2)\leftarrow PK(n-1)\leftarrow PK(n)
\label{statechain}
\end{equation}
that mirrors the inherent ordering of the time instants $n$.

It must be stressed that the possibility to order the states of knowledge $PK(n)$ has to be assumed, just like the existence of sequential time $n$ itself. The ordering in Eq. (\ref{statechain}) cannot be observed empirically by any subject, since the perceptions of any subject at any time $n$ is limited by definition to those objects that make up $PK(n)$. As noted in the beginning of this section, it is not possible, perception-wise, to transcend $PK(n)$ and observe the chain of states (\ref{statechain}) from outside in the manner shown in Fig. \ref{Figure10}(b). The assumption that there is such a chain can only be justified if it turns out to be helpful in the construction of a coherent model of the physical world and physical law.

Even if the ordering of events and states of potential knowledge $PK(n)$ is unambiguous in this transcendental sense, the perceived ordering of past events may become blurred as time passes, just as it was argued that it may be ambiguous from the subjective point of view whether a perceived object belongs to the present or to the past. If the perceived ordering of states of knowledge were indeed unambiguous, it would be possible to write

\begin{equation}\begin{array}{rcl}
M[PK(n-1)] & = & M[PKN(n-1)\cup M(PK(n-2))]\\
& = & M[PKN(n-1)]\cup M[M(PK(n-2))]
\end{array}
\label{preserveorder}
\end{equation}
for each $n$. If so, it would hold that

\begin{equation}\begin{array}{rcl}
PK(n) & = & PKN(n)\cup \\
& & M[PKN(n-1)]\cup \\
& & M^{2}[PKN(n-2)]\cup\\
& & \ldots\\
& & M^{m}[PKN(n-m)]\cup\\
& & \ldots
\end{array}
\label{perfectorder}
\end{equation}
and the ordering would be possible to read in state of knowledge $PK(n)$ from its decomposition into one set for each previous time, each marked with the exponent $m$ of the `memory hierarchy' $M^{m}$ that tells how far into history the associated memory of the present state $PKN(n-m)$ should be placed. However, the second equality in the relation (\ref{preserveorder}) is not necessarily fulfilled, so that the sequential time $m$ passed since a given event is not necessarily imprinted in the collective memory of all subjects at present time $n$.

\section{Relational time}
\label{relationaltime}

The role of sequential time $n$ is to represent the directed flow of time, which is seen as a fundamental form of appearance. Referring to the principle of ontological completeness discussed in section \ref{ocom}, a measure of temporal distances is also needed in the present approach, since the amount of time passed since a given event can indeed be subjectively felt and estimated.

The simplest way to define temporal distances would be to refer to the number $m$ introduced in the previous section, which counts the number of sequential time steps since a previous event. However, it is impossible for any subject to keep track of this number. Let the events $e^{k}_{i}$ and $e^{k}_{j}$ perceived by subject $k$ define the starting point and the endpoint of an attempt to find the number $m$ of temporal updates between them. After the event $e^{k}_{i}$ subject $k$ must keep track of all changes $e^{k'}$ of all objects perceived by all other subjects $k'$ in the universe. She cannot know the number of such objects. It may very well be infinite. The number $m$ of events $e^{k'}$ that happens in the universe between $e^{k}_{i}$ and $e^{k}_{j}$ is therefore unknowable to $k$ and possibly infinite. The impossibility to know the value of $m$ for any given subject $k$ disqualifies it as a universal physical measure of temporal distance, together with its possible divergence to infinity. To give it that role would violate the principle of ontological minimalism, discussed in section \ref{omin}.

For these reasons, \emph{relational time} $t$ is introduced as a measure of time. The quantity $t_{ll'}$ estimates the temporal distance between any two objects $O_{l}$ and $O_{l'}$ that are part of $PK(n)$, as illustrated in Fig. \ref{Figure10}(a). The value of $t_{ll'}$ may be uncertain, just like the value of any other attribute, to the extent that the temporal distance between two events may be completely unknown. The distance $t_{ll'}$ is defined by counting how many reference objects are placed temporally in between $O_{l}$ and $O_{l'}$. Of course, these reference objects correspond to the successive ticking of a clock.

Given the present epistemic approach to physics, this operational approach to the definition of distances between the values of a given attribute of two objects by counting the number of reference objects put in between them is the only reasonable one. Unsurprisingly, Kant used a similar line of reasoning when he connected the concept of \emph{magnitude} to intermediate reference objects, and noted that time is a necessary form of appearance in order to count them and thus make a measurement \cite{kant}.

\begin{quote}
\emph{No one can define the concept of magnitude in general except by something like this: That it is the determination of a thing through which it can be thought how many units are posited in it. Only this how-many-times is grounded on successive repetition, thus on time and the synthesis (of the homogeneous) in it.}
\end{quote}

There is no universal clock that all subjects can perceive. Therefore it must be allowed that different subjects $k$ and $k'$ use different clocks, and place different numbers of tickings between the same pair of events. In other words, a label $k$ should be attached to the distance $t_{ll'}^{k}$ and it should be allowed that $t_{ll'}^{k}\neq t_{ll'}^{k'}$. The knowledge about the value of $t_{ll'}^{k}$ becomes an object that is part of the potential knowledge $PK^{k}(n)$ of subject $k$ at time $n$, referring to the states $S_{O_{l}O_{l}}(n)$ and $S_{O_{l'}O_{l'}}(n)$ at the very \emph{same} time $n$. In this way the temporal distance $t_{ll'}^{k}$ becomes a \emph{perceived} attribute at a given time $n$ in contrast to the \emph{transcendent} sequential temporal distance $m$ that relates \emph{different} sequential times $n-m$ and $n$ in Eq. (\ref{perfectorder}).

The quantity $t_{ll'}^{k}$ is clearly a relational attribute, which needs more than one object to be defined. If subject $k$ chooses a spatio-temporal reference frame that contains an object assigned the role of an origin, a pseudo-internal attribute $t_{l}^{k}$ can be defined for each object $l$. The situation is analogous to that of spatial position $r_{l}^{k}$, which is relational in essence, but can be treated as an internal attribute of $l$ given a reference frame of other objects that defines a coordinate system. It should be noted that such spatio-temporal coordinate systems are defined at a given sequential time $n$. They can sometimes be identified with coordinate systems defined at previous sequential times, in the sense that the involved reference objects are identifiable in the sense of Eq. (\ref{id}).

The clock $c^{k}$ used by subject $k$ to measure $t_{ll'}^{k}$ can be described as an object $O_{c}^{k}$ that changes at a sequence of times $\{n,n+m,n+m',\ldots\}$, forming a corresponding sequence of events $\{e^{k}_{n},e^{k}_{n+m},e^{k}_{n+m'},\ldots\}$. Since it is impossible to determine the numbers $m,m',\ldots$, there is no inherent way to say whether the clock $c^{k}$ ticks at regular intervals or not. The only thing subject $k$ can do is to count the number of ticks $e^{k}_{n},e^{k}_{n+m},e^{k}_{n+m'},\ldots$ she perceives between the events $e^{k}_{i}$ and $e^{k}_{j}$, events that can be associated with a pair of object states $\{S_{O_{l}O_{l}}, S_{O_{l'}O_{l'}}\}$. The uncertainty of $t_{ll'}^{k}$ may stem either from the fact that the memories of the past ticks may be imperfect, as expressed by Eq. (\ref{sloppymemory}), or from the use of a clock with poor resolution, which places only a few ticks between a typical pairs of events. A measure of the uncertainty is defined from a comparison with a reference clock with better resolution, one which places several ticks between each adjacent pairs of ticks of the cruder clock.

Even though each value $t_{ll'}^{k}$ may be more or less uncertain, its role as a temporal measure makes it possible to state a set of relations that selected sets of values $\{t_{ll'}^{k}\}$ must fulfill. In the language of quantum mechanics, they represent entanglement between the states of the involved objects. For each subject $k$ and each set of objects $\{O_{1},O_{2},\ldots,O_{N}\}$ with arbitrary labeling, it holds that

\begin{equation}
0=t_{12}^{k}+t_{23}^{k}+\ldots+t_{N-1,N}^{k}+t_{N1}^{k}.
\label{circlezero}
\end{equation}
In the case $N=2$ the relation $0=t_{ll'}^{k}+t_{l'l}^{k}$ is fulfilled for any two objects $O_{l}$ and $O_{l'}$, expressing the directed nature of time. The case $N=3$ is illustrated by the cyclic set of temporal distances $\{t_{23}^{k},t_{34}^{k},t_{42}^{k}\}$ shown in in Fig. \ref{Figure10}(a). In addition, it must be required that

\begin{equation}
t_{ll'}^{k}=0
\end{equation}
for each subject $k$ whenever $Pr=1$ for both objects $O_{l}$ and $O_{l'}$, as illustrated in Fig. \ref{Figure10}(a) for $t_{45}^{k}$.

\begin{figure}[tp]
\begin{center}
\includegraphics[width=80mm,clip=true]{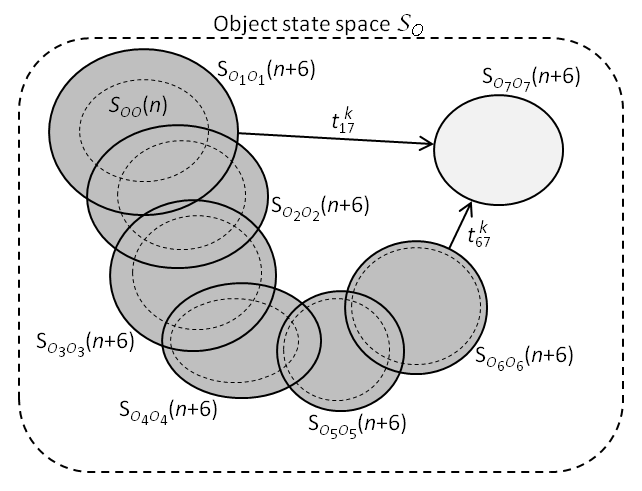}
\end{center}
\caption{The memories at time $n+6$ of the sequence of states shown in Fig. \ref {Figure9}. Each such memory is an object $O_{m}$ with state $S_{O_{m}O_{m}}(n+6)$ that refers to the same object $O$ at time $n+m-1$ with state $S_{OO}(n+m-1)$. At the given sequential time $n+6$, relational temporal distances $t_{ll'}^{k}$ are defined between any two objects $O_{l}$ and $O_{l'}$.}
\label{Figure11}
\end{figure}

The relation between sequential time $n$ and relational time $t$ is highlighted by a comparison between Fig. \ref{Figure9} and Fig. \ref{Figure11}. All object states shown in Fig. \ref{Figure11} are defined at the same time $n+6$ just after object $O$ in Fig. \ref{Figure9} has undergone a knowable change. This means that all other object states shown in Fig. \ref{Figure11} correspond to memories of past states of $O$. For example, it holds that $S_{O_{1}O_{1}}(n+6)=M^{6}(S_{OO}(n))$ in the notation used in Eq. (\ref{perfectorder}). The fact that memories may be imperfect according to Eq. (\ref{sloppymemory}) is illustrated by the fact that remembered object states like $S_{O_{1}O_{1}}(n+6)$ (solid lines) contain the corresponding original object states like $S_{OO}(n)$ (dashed lines) as subsets. In this context the presentness attribute $Pr$ is disregarded.

\subsection{Discrete time and continuous space}
\label{discont}
In this section, it is argued that relational time $t$ is inherently discrete, since it is associated with discrete sequential time $n$. This quality of $t$ is contrasted to that of the spatial distance $x$, which is potentially continuous, according to a simple argument presented below. In the next section it will be argued that the continuous nature of $x$ makes it possible to consider $t$ continuous as well in certain situations, namely when $t$ refers to a quasiobject whose temporal position is deduced from it spatial position and a given velocity.

The argument starts with a question: What value should be assigned to the temporal distance $t_{67}^{k}$ between the present object state $S_{O_{7}O_{7}}(n+6)$ in Fig. \ref{Figure11} and the memory $S_{O_{6}O_{6}}(n+6)$ of the immediately preceding state $S_{OO}(n+5)$ of the same object $O$ just before it knowably changed? The change corresponds to a single event $e^{k}$ potentially perceived by some subject $k$, according to Eq. (\ref{event}). Therefore there is no set of other events perceived by $k$, no clock ticks, which can be placed in between these two object states, whose number would define $t_{67}^{k}$. Nevertheless, the assignment $t_{67}^{k}=0$ is improper, since the absence of a temporal distance between a state before and after an event would mean that there are no temporal distances at all between any pair of object states. The only reasonable choice is to set $t_{67}^{k}=1$, reflecting the fact that events define the passage of time, and temporal distances $t_{ll'}^{k}$ are defined by counting the number of events that subject $k$ can place between the perceptions of the object states $S_{O_{l}O_{l}}$ and $S_{O_{l'}O_{l'}}$.

Therefore, it seems a minimum temporal distance $t_{\min}=1$ must be introduced, which applies to the temporal distance between the memories of any two subsequent events, defining two subsequent temporal updates $n\rightarrow n+1$ and $n+1\rightarrow n+2$. To each of these events a corresponding unit increment of relational time $t$ is assigned according to the preceding paragraph, but between them nothing happens that could motivate any such increment. Actually, the words "between them" lack epistemic meaning.

\begin{figure}[tp]
\begin{center}
\includegraphics[width=80mm,clip=true]{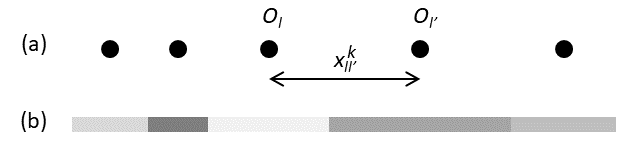}
\end{center}
\caption{The very notion that two objects $O_{l}$ and $O_{l'}$ can be spatially distinguished implies that there is something in between them. This "something" corresponds to another object $O_{b}$, showing that there cannot be any closest possible pair of distinct objects, and no smallest non-zero spatial distance $x_{ll'}^{k}$. This is true regardless if $O_{l}$ and $O_{l'}$ are perceived as (a) spatially separated, (b) spatially extended, or both.}
\label{Figure12}
\end{figure}

The situation is different when it comes to spatial distances $x_{ll'}^{k}$ between two objects $O_{l}$ and $O_{l'}$ with states $S_{O_{l}O_{l}}$ and $S_{O_{l'}O_{l'}}$, as illustrated in Fig. \ref{Figure12}. Whenever two such objects are perceived as spatially separated, so that the assignment $x_{ll'}^{k}>0$ has to be made, at least one more object that is placed between them is also perceived.

This statement may need some explanation. Suppose that no object is explicitly seen between $O_{l}$ and $O_{l'}$, making them a candidate of a closest possible pair. Then there are two main alternatives. They may be separated by a perceived void, as shown in Fig. \ref{Figure12}(a). Then the void itself becomes the intermediate object $O_{b}$, since in this study an object is broadly defined as an element of perception.

Alternatively, $O_{l}$ and $O_{l'}$ may be perceived as spatially extended and touching each other, where an internal attribute like brightness changes value at the interface, as shown in Fig. \ref{Figure12}(b), so that the two objects become distinguishable and worthy of separate labels. The assumed spatial extension of $O_{l}$ and $O_{l'}$ means that $x_{ll'}^{k}$ must be defined between their centers of mass, or by a similar criterion. By definition of spatial extension, smaller parts $O_{b}$ can then be distinguished between the two centers of mass, between which distances smaller than $x_{ll'}^{k}$ must be defined. One such intermediate part, or object, is the interface between $O_{l}$ and $O_{l'}$, where the brightness or some other attribute changes value.

Of course, two spatially separated objects may be separated by a perceived void as well as being spatially extended. Then both the above arguments show that there is a third object implied between them. In any case, it is clear that there cannot be any smallest spatial distance, from the chosen cognitive point of view. Between any two objects, a third is implied. Therefore spatial distance should be treated as a potentially continuous relational attribute in the present approach to physics.

\subsection{Continuous deduced time}
\label{contded}

\begin{figure}[tp]
\begin{center}
\includegraphics[width=80mm,clip=true]{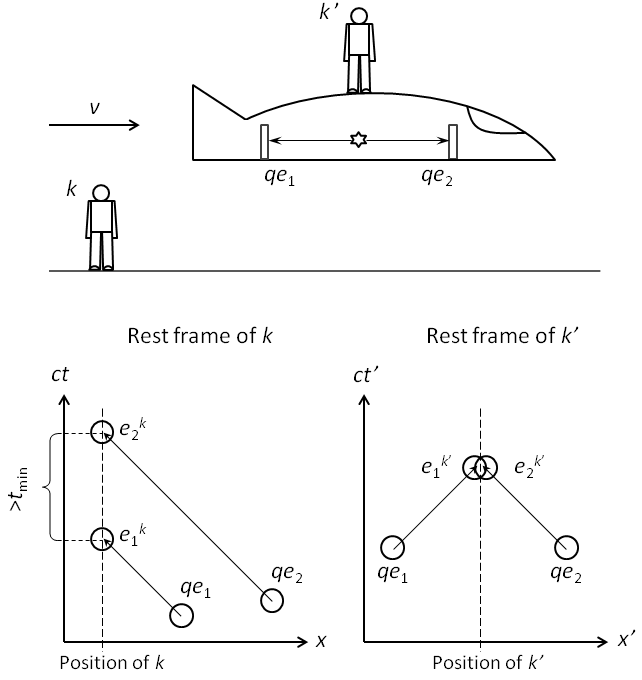}
\end{center}
\caption{The time difference between perceived events is either greater than $t_{\min}$ (like that between $e_{1}^{k}$ and $e_{2}^{k}$), or zero (like that between $e_{1}^{k'}$ and $e_{2}^{k'}$). All subjects agree on these statements. The fact that the temporal ordering of the quasievents $qe_{1}$ and $qe_{2}$ is ambiguous has no primary importance in the present epistemic approach. The temporal distance between two quasievents may take any real value.} 
\label{Figure13}
\end{figure}

It was argued in section \ref{sequential} that if perceived events are to be embedded in space-time, they have to be located along the world lines of the bodies of the perceiving subjects. The discussion in section \ref{discont} implies that the proper (relational) time passed between a pair of events along any such world line is either zero or larger than some discrete temporal unit, due to the inherent discreteness of a series of events. Figure \ref{Figure13} illustrates the fact that it is nevertheless possible to preserve the notion of a continuous space-time, thanks to the continuity of spatial distances according to the discussion in section \ref{discont}.

To justify this statement, note first that the temporal distance $t_{12}^{k}$ measured by subject $k$ in her own rest frame between any two perceived events $e_{1}^{k}$ and $e_{2}^{k}$ always fulfills $t_{12}^{k}=0$ or $t_{12}^{k}\geq t_{\min}$, where the condition $t_{\min}=1$ is dropped to allow for arbitrary units. The pair of events $e_{1}^{k}$ and $e_{2}^{k}$ correspond to a pair of perceived objects $O_{1}$ and $O_{2}$.

Subject $k$ may properly interpret the perceived objects as corresponding to two objects that are placed at some distance from $k$, the information about which arrives to $k$ at the speed of light. In the vocabulary of section \ref{idquasi}, these correspond to two quasiobjects, which may be denoted $QO_{1}$ and $QO_{2}$. In Fig. \ref{Figure13} two such two quasiobjects correspond to the events that a flash of light emitted from the middle of a space ship is reflected at a front and a rear mirror, respectively. These events are not perceived by anybody, and may thus be called two \emph{quasievents} $qe_{1}$ and $qe_{2}$.

Since spatial distances are treated as continuous, each of the two quasievents $qe_{1}$ and $qe_{2}$ may be located anywhere on the light rays arriving at the world lines of subject $k$ or $k'$, so that any real temporal coordinate $t$ or $t'$ can be assigned to them. The deduced temporal distance $t_{Q12}^{k}$ between $qe_{1}$ and $qe_{2}$ can be arbitrary small but non-zero if their separation is space-like, even though $t_{12}^{k}\geq t_{\min}$.

\subsection{Relativity of simultaneity}
\label{revsim}

The assumption that different subjects may perceive the same objects and events sometimes corresponds to the fact that their differing observations can be deduced to correspond to the same quasiobjects or quasievents, as discussed in section \ref{idquasi}. This means that these quasiobjects and their changes, as deduced by different subjects, are not distinguishable in sense discussed in relation to Fig. \ref{Figure4}(b), and thus must be identified from an epistemic point of view.

An example is given in Fig. \ref{Figure13}, where a subject $k'$ perceives a pair of events $e_{1}^{k'}$ and $e_{2}^{k'}$ that correspond to the same quasievents $qe_{1}$ and $qe_{2}$ as those perceived by $k$. Subject $k$ perceives event $e_{1}^{k}$ before event $e_{2}^{k}$ and deduces that $t_{Q12}^{k}>0$. In contrast, $k'$ perceives $e_{1}^{k'}$ and $e_{2}^{k'}$ as simultaneous, deducing that $t_{12}^{k'}=0$. This circumstance expresses the relativity of simultaneity.

On the other hand, the temporal ordering of the corresponding directly perceived events $e_{1}^{k}$, $e_{2}^{k}$, $e_{1}^{k'}$, $e_{2}^{k'}$ shown in Fig. \ref{Figure13} is assumed to be well-defined and universal in terms of sequential time $n$. For example, it may hold that

\begin{equation}
n[e_{1}^{k}]<n[e_{1}^{k'}]=n[e_{2}^{k'}]<n[e_{2}^{k}],
\end{equation}
where $n[e]$ is the sequential time at which the observed object $O_{1}$ that defines event $e$ changes according to Eq. (\ref{event2}).

It is seen that from the present epistemic perspective the relativity of simultaneity applies to quasiobjects or quasievents with \emph{deduced} spatio-temporal location $(x,t)$, whereas the temporal ordering of \emph{perceived} events by means of sequential time $n$ may still be unambiguous and universal according to the discussion in section \ref{sequential}. It is also seen that it is possible from the epistemic perspective to immerse all objects known at a given sequential time $n$ in a Minkowski space-time, provided that the roles of objects $O$ and quasiobjects $QO$ are distinguished.

\begin{figure}[tp]
\begin{center}
\includegraphics[width=80mm,clip=true]{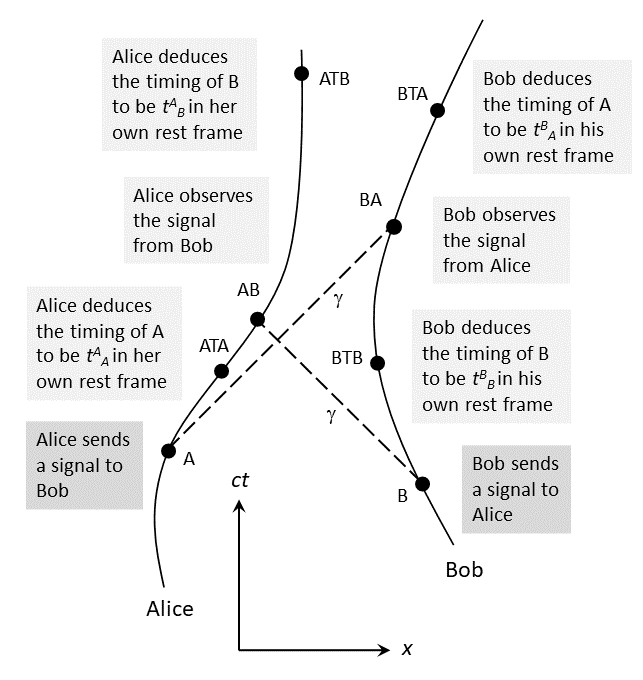}
\end{center}
\caption{Alice and Bob may disagree on who sent a signal first, due to the relativity of simultaneity when deduced quasievents that are not situated on the world line of the observer are involved. Nevertheless, it may be assumed without contradiction that the temporal ordering in terms of sequential time $n$ of the eight depicted events is universal. See text for further explanation.} 
\label{Figure14}
\end{figure}

Figure \ref{Figure14} tries to clarify these points further. Suppose that Alice and Bob are traversing the universe far from each other in two spaceships. At about the same time they suddenly miss each other and send a signal to their friend. These are the events A and B in Fig. \ref{Figure14}. They look at their clocks to memorize when they sent their signal in order to know when they can expect a reply, given the limited speed of light. The checking corresponds to the events ATA and BTB, and they find the times $t_{A}^{A}$ and $t_{B}^{B}$, respectively. Each of them is happy to get a response from their friend earlier than expected, corresponding to the events AB and BA. They check their clocks to conclude that the response arrived so soon that their friend must have sent it before the signal they sent themselves arrived. These are the events ATB and BTA, and the deduced times when their friend sent the signal are $t_{B}^{A}$ and $t_{A}^{B}$, respectively. Both Alice and Bob are happy to realize that their friend must have thought of them and tried to contact them by themselves, rather than just replying to a signal.

Clearly, this means that the events A and B have space-like separation. But which signal was sent first? Relativistically, the answer is ambiguous. Alice may deduce that

\begin{equation}
t_{A}^{A}-t_{B}^{A}>0,
\end{equation}
concluding that Bob thought of her first, and Bob may deduce that

\begin{equation}
t_{A}^{B}-t_{B}^{B}<0,
\end{equation}
concluding that Alice thought of him first. It is argued however, that the question has a definitive answer in terms of sequential time $n$. The temporal ordering of the involved events may be

\begin{equation}
n_{B}<n_{A}<n_{BTB}<n_{ATA}<n_{AB}<n_{BA}<n_{ATB}<n_{BTA},
\label{sevents}
\end{equation}
meaning that Bob actually signaled Alice before Alice signaled Bob. If you believe in thought transference, you may imagine that Bob inspired Alice to send a signal.

It is important to note that the sequence of events in Eq. (\ref{sevents}) is not embedded in any space-time \emph{per se}. Therefore there is no contradiction in the claim that their ordering is well-defined, despite the relativity of simultaneity. Alice constructs the space-time in which she concludes that Bob signaled her first at sequential time $n_{ATB}$, and Bob constructs the corresponding space-time in which he concludes that Alice signaled him first at sequential time $n_{BTA}$. The space-time shown in Fig. \ref{Figure14} corresponds to a construction by a third party observer at some sequential time $n>n_{BTA}$, at which information about all the depicted events has had time to reach her at the speed of light.

\section{The evolution parameter}
\label{evolp}

The world seemingly changes gradually. The shorter time passed, the smaller changes of the perceived objects. This is not self-evident in the present picture of time. The subjective ability to order events temporally is simply assumed, regardless the similarity or dissimilarity of subsequent object states $S_{OO}$. However, a gradual change of the object states is necessary in order to use Eq. (\ref{id}) to define identifiable objects, and to be able to speak about the trajectory of a given object, as illustrated in Fig. \ref{Figure9}. Such identifiability is necessary in order to say that the world perceived at time $n+m$ is \emph{the same} as that perceived at some previous time $n$, according to the discussion in section \ref{evolstates}. Therefore, it is essential that the evolution operators $u_{1}$ and $u_{O1}$ introduced in Eqs. (\ref{sevolution}) and (\ref{soevolution}), respectively, express such a gradual change.

More than that, the overlapping subsequent object states shown in Fig. \ref{Figure9} makes the evolution of an identifiable object $O$ seamless. It becomes impossible to tell two subsequent object states apart, but it may nevertheless be possible to perceive that the state of the object has changed after a longer period of time. For example, it is seen in Fig. \ref{Figure9} that the states $S_{OO}(n)$ and $S_{OO}(n+5)$ do not overlap, corresponding to the fact that they are possible to distinguish.

\begin{figure}[tp]
\begin{center}
\includegraphics[width=80mm,clip=true]{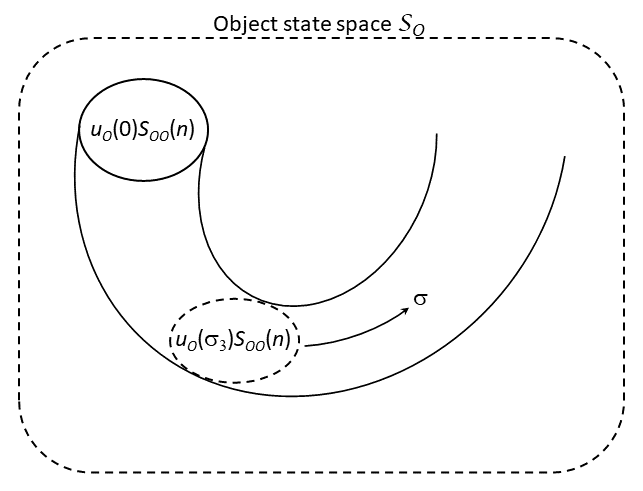}
\end{center}
\caption{The apparent fact that physical evolution is gradual can be expressed as the possibility to find a representation $u_{O}(\sigma)$ of the stepwise evolution operator $u_{Om}$ that depends continuously on an evolution parameter $\sigma$. Compare Fig. \ref{Figure9}.}
\label{Figure15}
\end{figure}

These facts make it natural to model the evolution of $O$ as if it follows a continuous trajectory. In this spirit, a continuous operator $u_{O}(\sigma)$ may be introduced, which depends on an evolution parameter $\sigma\in\mathbf{\mathbb{R}}$ such that

\begin{equation}
u_{O}(\sigma_{m})\equiv u_{Om},
\label{sigmadef}
\end{equation}
where $\sigma_{1}<\sigma_{2}<\sigma_{3}<\ldots$ and $u_{Om}$ is the $m$-step evolution operator

\begin{equation}
u_{Om}\equiv u_{O1}[u_{m-1}S(n)]u_{O1}[u_{m-2}S(n)]\ldots u_{O1}[u_{1}S(n)]u_{O1}[S(n)],
\label{uomdef}
\end{equation}
where the $1$-step evolution operator $u_{O1}$ is defined in Eq. (\ref{soevolution}), and $u_{m}\equiv(u_{1})^{m}$. The operator $u_{Om}$ reveals all that can be predicted about the object state $S_{OO}(n+m)$ at time $n$. The discrete evolution of object $O$ in Fig. \ref{Figure9} is expressed in Fig. \ref{Figure15} in a continuous fashion by means of the evolution parameter $\sigma$ as an argument of the evolution operator, where the convention $u_{O}(0)=I$ is used. 

\begin{figure}[tp]
\begin{center}
\includegraphics[width=80mm,clip=true]{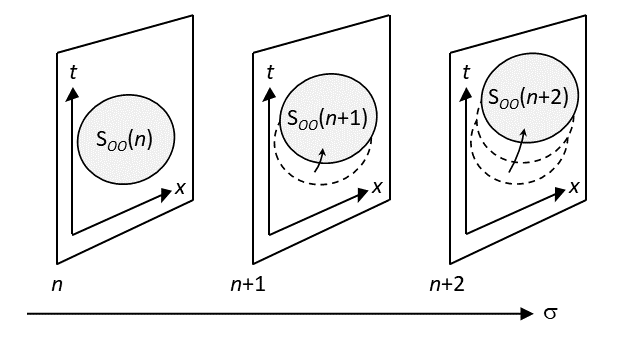}
\end{center}
\caption{The relation between the evolution parameter $\sigma$, sequential time $n$, and relational time $t$. There is an entire space-time spanned by the spatial axis $x$ and the temporal axis $t$ associated with each sequential time $n$, onto which the state $S_{OO}(n)$ of object $O$ can be projected. The seamless evolution of $S_{OO}(n)$ as $n$ increases can be described using the continuous parameter $\sigma$. At each time $n$ there are memories of past states of $O$, which are shown as states with dashed boundaries. They make it possible to define a trajectory $x(t)$ of $O$ at each time $n$, as indicated by curved arrows.}
\label{Figure16}
\end{figure}

The relation between the three temporal quantities $\sigma$, $n$ and $t$ is illustrated schematically in Fig. \ref{Figure16}. It was argued in section \ref{presentpast} that sequential time $n$ is a transcendental quantity rather than an observable attribute. The same goes for $\sigma$ since it is defined via the flow of sequential time according to Eq. (\ref{sigmadef}). This means that its value is unknowable, just like the value $n$ of sequential time, and it therefore lacks physical meaning. Any invertible change of variables $\sigma'=f(\sigma)$ produces an equally valid evolution parameter $\sigma'$.

The evolution parameter $\sigma$ is useful to answer the question what the object $O$ will look like when it is examined the next time after an examination at time $n$, depending on how long the observer is waiting before the second examination. This waiting time is defined by the number $m$ of changes of \emph{other} objects $O'$ that are observed before she looks at $O$ again, and $m$ may in principle be any positive integer. The adjustable $m$ and the seamless evolution of the object state make it meaningful to use $\sigma$ to express continuous evolution equations applying to any object $O$.

In contrast, it is not meaningful to express such a continuous evolution equation that applies to the entire world. All that can be said about $S(n+1)$ is a function of $S(n)$. These two states do not overlap, as illustrated in Fig. \ref{Figure5}(a), so that there is no sense in which the evolution of the world as a whole is seamless, from the cognitive viewpoint used in this paper. Further, observers cannot choose to wait an arbitrary amount $m$ of sequential time after time $n$ until the world is observed the next time, since $n+1$ corresponds by definition to the next time someone looks at it. The discrete mapping in Eq. (\ref{sevolution}) defined by the evolution operator $u_{1}$ is sufficient.

\section{Discussion}

The idea to separate two aspects of time in a manner similar to that outlined here is not new. Bernard Carr argues that the perceived flow of time should be modelled as a 'mental time' that can be separated from the 'physical time' that is part of ordinary space-time \cite{carr}. Each 'now' corresponds to a point along the mental temporal axis associated with an entire space-time, in which the future trajectories are uncertain, whereas the past trajectories are well-defined. Carr traces this idea back to Charlie Dunbar Broad \cite{broad}.

Despite the similarities, the above perspective differs in some respects from the one adopted in this paper. Here, the two aspects of time are both 'mental' in the sense that they correspond to forms of appearances, namely the perceived flow of time and the possibility to measure temporal distances. At the same time, they are both 'physical' in the sense that proper physical models are assumed to be representations of the perceived or perceivable world, as illustrated in Fig. \ref{Figure2}.   

More precisely, a universal ordering of events is constructed from an cognitive starting point, where an event corresponds to a change of the potential knowledge of some observer. This directed sequence defines sequential time $n$. At each time $n$ there is knowledge about a set of present and past events, as well as the spatial and temporal relations between them, quantified by $x$ and $t$, respectively. In other words, there is an entire space-time associated to each $n$.

It is possible to associate a continuous evolution parameter $\sigma$ to the discrete flow of time described by $n$. The quantity $\sigma$ can be used to parametrize the smooth evolution of identifiable objects that emerges from the analysis. The numerical value of $\sigma$ can be precisely defined, but lacks physical meaning.

The numerical values of $t$, on the other hand, has physical meaning as temporal distances that are measured and known by some observer at a given time $n$. As such, the knowledge about $t$ may be incomplete, just like the knowledge about spatial distances $x$, or any other relational or internal attribute of an object. From the present epistemic perspective, uncertainties $\Delta t$ and $\Delta x$ must therefore be allowed as fundamental ingredients in the physical formalism. They can be interpreted as Heisenberg uncertainties. The space-time that emerges from the present analysis thus gets quantum mechanical qualities. It becomes possible to speak about superpositions of space-times  at a given sequential time $n$, each of which is consistent with the incomplete potential knowledge available at that time.

It is natural to associate relational time $t$ to the aspect of time used in relativity theory, where it is treated as an observable, and to associate the evolution parameter $\sigma$ to the aspect of time used to express quantum mechanical evolution equations, where it is treated as a precisely defined evolution parameter, without any associated Heisenberg uncertainty.

If such an association is indeed possible, the present approach may indicate a way forward in order to solve the \emph{problem of time}. Such a solution would not be to eliminate time, or to choose either the relativistic or the quantum theoretical perspective on time, but to incorporate them both into the physical formalism. In an accompanying paper, a relativistically covariant formalism of that kind is introduced \cite{ostborn2}. The starting point will be the epistemic physical object states $S_{OO}(n)$ and their evolution, as described in sections \ref{states} and \ref{evolstates}.

The choice to express the evolution of physical states in terms of sequential time $n$, or the associated evolution parameter $\sigma$, can be motivated as follows. Relational time $t$ encodes the temporal relations known at a certain time $n$ between present events and memories of past events, or between different memories of past events. To vary sequential time $n$ means to transcend the knowledge about the world at a certain time. Therefore $n$ is adequate to express the physical laws responsible for the evolution of the physical state from one moment of time to the next. This is conceptually coherent since physical law by definition transcends the individual physical states it applies to.

A core idea of this paper is to expand the fundamental physical description of the `now' so that it contains information not only about the state of the world at that given moment $n$, but also partial information about the past. Such an expansion is justifiable only if an epistemic approach to physics is adopted, in which the present physical state of the world corresponds to the present knowledge about the world. That knowledge contains information about the past via memories and records. In contrast, in a realistic model of the world, the information about the past in the form of memories and records is just a function of the present physical state of the brain \cite{hartle}. The only relevant aspect of the records will be their present physical state. Thus, memories and records become secondary from a realistic point of view, and the past does not become a necessary part of the description of the present.

From the subjective point of view, the expanded notion of the `now' is the more natural one. Suppose that you listen to music. The appreciation of harmonies, and the emotional response they give rise to in the present, depend crucially on memories of sounds in the immediate past, to the extent that the music would cease to exist without these memories. That is, each present state of the listener contains both the present and the past in a crucial way; each fleeting `now' encoded by $n$ can be unfolded to an entire temporal axis $t$. At the formal level of physical description, the very perception of a sound at a given moment relies on sensory recording during an extended period of time, since such a temporal interval is needed to define the frequencies that determine the sound that we hear at a given moment. 

To conclude, allow some philosophical digressions. An epistemic approach to physics, such as the present one, must nevertheless possess an ontology that refers to a transcendental aspect of the world, just like all other approaches to physics. Physical law itself transcends the set of perceptions that are used to deduce it. It has been argued that sequential time $n$ is transcendental in the sense that we cannot perceive all instants along the sequential time axis. We are stuck at a particular point in time $\hat{n}$ that we call `now', from which we remember past events or read records about them.

\begin{figure}[tp]
\begin{center}
\includegraphics[width=80mm,clip=true]{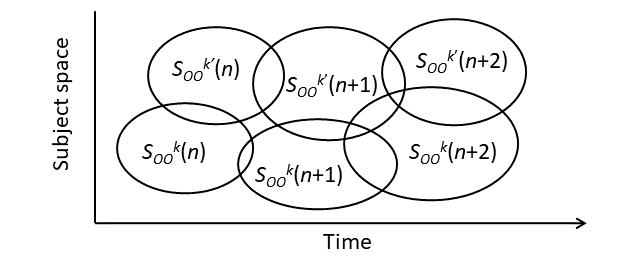}
\end{center}
\caption{Overlapping object states $S_{OO}$ weave a single world that preserves its identity as time passes and contains several subjects. Here, a single identifiable object $O$ is perceived at times $n$, $n+1$ and $n+2$ by two subjects $k$ and $k'$. Compare Figs. \ref{Figure1} and \ref{Figure3} with regard to different subjects perceiving the same object, and Fig. \ref{Figure9} concerning the definition of an identifiable object.}
\label{Figure17}
\end{figure}

To accept the existence of other times $n$ requires a leap of faith. Such a leap is necessary in order to arrange these time instants sequentially, and to use this arrangement as a basis of a physical model of evolution, like in the present paper. A similar leap of faith is needed in order to believe in other subjects. Just as past times can only be accessed indirectly via perceptions of memories at the present time, other subjects can only be accessed indirectly via our own perceptions of bodies of other subjects. Just as a link is postulated in Fig. \ref{Figure10}(b) between a memory and a past event, a link has to postulated between a body of someone and that subject. We speak about a memory \emph{of} something and the body \emph{of} someone, where the the word \emph{of} expresses these links. Further, from the subjective point of view, the structure of time and mind is asymmetric in the same way: there is a vantage point from which the entire field is seen. The analogy is manifest in language, via a certain congruence between the triplets of words (now, then, time) and (I, you, mind).

From the conventional point of view, these asymmetries and leaps of faith are just curiosities of perception with no relevance to physics. Applying Kant's Copernican revolution to physics, however, they become fundamental forms of appearances on which physical models should be built. Figure \ref{Figure17} illustrates, for example, how overlapping object states weave together different subjects and different times into a common physical world, again illustrating the similarity between the temporal and individual degrees of freedom. There is a difference, though: the temporal degree of freedom is directed, whereas the individual degree of freedom is not. There is no inherent hierarchy among subjects, as far as physics concerns. For instance, the reference frame of each observer is equally valid in relativity theory, in a perfectly democratic fashion.

The leaps of faith along the temporal and individual degrees of freedom are essential in the cognitive physical model built in this paper. This is so, in particular, in the construction of sequential time $n$. No single individual can ever perceive this universal temporal chain of events, even though each link in the chain is an element of perception. On the other hand, it is equally essential in the proposed model that no leap of faith is allowed for the purpose of introducing objects, or attributes of objects, that cannot in principle be perceived by anyone, that are beyond perception in general. Again, this is Kant's Copernican revolution applied to physics.

The epistemic approach to physics outlined in this paper is not proposed on its philosophical merits, but is presented as a perspective that is potentially fruitful for physics. It is considered to have merit if it can be used to increase the degree of coherence in physical models, or in the understanding of these models. The approach has even more merit if it can provide new physical predictions that can be tested empirically. This paper does not go all the way to fulfill either of these criteria, but provides a point of departure for attempts to do so when it comes to the treatment of time. One step further along that road is attempted in the accompanying paper \cite{ostborn2}.

\end{document}